\documentclass[twocolumn,showpacs,preprintnumbers,amsmath,amssymb]{revtex4}
\usepackage{amsmath,epsfig,subfigure}
\usepackage{graphicx}
\usepackage{dcolumn}
\usepackage{bm}
\usepackage{graphicx}

\newcount\EquaNo \EquaNo=0
\def\newEq#1{\advance\EquaNo by 1 #1=\EquaNo}
\newcount\TablNo \TablNo=0
\def\newTabl#1{\advance\TablNo by 1 #1=\TablNo}
\newcount\FigNo \FigNo=0
\def\newFig#1{\advance\FigNo by 1 #1=\FigNo}
\newcount\ChapNo \ChapNo=0
\def\newCh#1{\advance\ChapNo by 1 #1=\ChapNo}
\def\expec#1{\big\langle{#1}\big\rangle}
\begin {document}

\title { Steady-State Properties of Single-File Systems with Conversion}

\author{S.V. Nedea}
\altaffiliation{ Department of Mathematics and Computing Science, Eindhoven University of Technology, P.O. Box 513, 5600 MB Eindhoven, The Netherlands.}
\email{silvia@win.tue.nl}
\author{A.P.J. Jansen}
\altaffiliation{
 Department of Chemical Engineering, Eindhoven University of Technology, P.O. Box 513, 5600 MB Eindhoven, The Netherlands.
}
\author{J.J. Lukkien}
\altaffiliation{ Department of Mathematics and Computer Science,
Eindhoven University of Technology, P.O. Box 513, 5600 MB
Eindhoven, The Netherlands.}
\author{P.A.J. Hilbers}
\altaffiliation{ Department of Biomedical Engineering, Eindhoven
University of Technology, P.O. Box 513, 5600 MB Eindhoven, The
Netherlands.}

\date{\today}

\begin {abstract}

 We have used Monte-Carlo methods and analytical techniques to
investigate the influence of the characteristic parameters, such as pipe
length, diffusion, adsorption, desorption and reaction rate constants on the
steady-state properties of Single-File Systems with a reaction.
  We looked at cases when all the sites are reactive and when only some of them
are reactive. Comparisons between Mean-Field predictions and Monte Carlo
simulations for the occupancy profiles and reactivity are made. Substantial
differences between Mean-Field and the simulations are found when rates of diffusion
are high. Mean-Field results only include Single-File behavior by changing the
diffusion rate constant, but it effectively allows passing of particles.
 Reactivity converges to a limit value if more reactive sites are added: sites
in the middle of the system have little or no effect on the kinetics. Occupancy
profiles show approximately exponential behavior from the ends to the middle of the system.

\end{abstract}

\maketitle

\section { Introduction }

  Molecular sieves are crystalline materials with open framework structures.
Of the almost two billion pounds of molecular sieves produced in the last
decade, 1.4 billion pounds were used in detergents, 160 millions pounds as
catalysts and about 70 millions pounds as adsorbents or desiccants.~\cite{Lloyd}

  Zeolites represent a large fraction of known molecular sieves. These are
all aluminosilicates with well-defined pore structures. In these
crystalline materials, the metal atoms (classically, silicon or aluminum)
are surrounded by four oxygen anions to form an approximate tetrahedron.
 These tetrahedra then stack in regular arrays such that channels and cages
are formed. The possible ways for the stacking to occur is virtually unlimited, and
hundreds of unique structures are known.~\cite{meier}

 The channels (or pores) of zeolites generally have cross section somewhat
larger than a benzene molecule. Some zeolites have one-dimensional channels parallel to one
another and no connecting cages large enough for guest molecules to cross
from one channel to the next. The one-dimensional nature leads to extraordinary
effects on the kinetic properties of these materials. Molecules move in a concerted
fashion, as they are unable to pass each other in the channels. These
structures are modeled by one-dimensional systems called Single-File Systems
where particles are not able to pass each other. A particle can only move to an
adjacent site if that site is not occupied.

  This process of Single-File diffusion has different characteristics than
ordinary diffusion which affects the nature of both transport and conversion by
chemical reactions. For Single-File diffusion, the mean-square displacement
of a particular particle is proportional to the square-root of time
    $$\langle r^2  \rangle = 2Ft^{1\over{2}}$$
where F is the Single-File mobility.~\cite{Sholl} This is in contrast to normal diffusion,
where mean-square displacement is directly proportional to time.
  A variety of approaches have been used to describe the movement of the
particles in Single-File Systems, most of them concentrated on the role of
the Single-File diffusion process.

 Molecular Dynamic(MD) studies of diffusion in zeolites have become increasingly
popular with the advent of powerful computers and improved algorithms.
 In a MD simulation the movement is calculated by computing all forces
exerted upon the individual particles. MD results
have been found to match experimental observations of Single-File diffusion
for systems with one type of molecule without conversion and with very short
pores.~\cite{keffer1,keffer2,keffer3,keffer4} Because a molecule can move
to the right or to the left neighboring site only if this site is free, MD simulations
under heavy load circumstances require a high computational effort for particles
that hardly move. However, the level of detail provided by MD simulations is
not always necessary.

 Thus, deterministic models are used also but they are mainly focused on dynamic and
steady-state information of short pore systems.~\cite{tsikoyannis,rodenbeck,okino}
 Several researchers~\cite{petzold,beijeren,hahn} used a stochastic
approach, i.e., Dynamic Monte Carlo(DMC), to determine the properties of Single-File Systems.
 In DMC reactions can be included. The rates of the reactions determine the probability with which
different configurations are generated and how fast (at what moment in time)
new configurations are generated. The most severe limitation of the DMC method arises when the
reaction types in a model can be partitioned into 2 classes with vastly
different reaction rates. In this case, extremly large amounts of computer
time are required to simulate a reasonable number of chemical reactions.
However, in general the system can be simulated for much longer times than
with MD.

 All the previous references put the emphasis on the transport properties of
adsorbed molecules  as the important factor in separation and reaction
processes that take place within zeolites and other shape-selective
microporous catalysts.
  R{$\ddot {\rm o}$}denbeck and K{$\ddot{\rm a}$}rger~\cite{rodenbeck} solved numerically the principal dependence of
steady-state properties such as concentration profiles and the residence
time distribution of the particles, on the system parameters for sufficiently
short pores.
 In multiple papers, Auerbach et al.~\cite{nelson,metiu} used Dynamic Monte Carlo to show
different predictions about Single-File transport and direct measurements of
intercage hopping ion strongly adsorbing quest-zeolite systems.
  Saravanan and Auerbach ~\cite{saravanan1, saravanan2} studied a lattice model
of self-diffusion in nanopores, to explore the influence of loading, temperature
and adsorbate coupling on benzene self-diffusion in Na-X and Na-Y zeolites.
 They applied Mean-Field(MF) approximation for a wide set of parameters, and derived an
analytical diffusion theory to calculate diffusion coefficients for various
loadings at fixed temperature, denoted as "diffusion isotherms". They found
that diffusion isotherms can be segregated into subcritical and
supercritical regimes, depending upon the system temperature relative to the
critical temperature of the confined fluid. Supercritical systems exhibit three
characteristic loading dependencies of diffusion depending on the degree of
degeneracy of the lattice while the subcritical diffusion systems are dominated
by cluster formation.
  Coppens and Bell~\cite{coppens1,coppens2,coppens3} studied the influence of occupancy and
pore network topology on tracer and transport diffusion in zeolites.
They found that diffusion in zeolites strongly depends on the pore network
topology and on the types and fractions of the different adsorption sites.
MF calculations can quickly estimate the diffusivity, although large
deviations from the DMC values occur when long-time correlations are present
at higher occupancies, when the site distribution is strongly heterogeneous
and the connectivity of the network low.

 Few researchers included also reactivity in Single-File Systems.
Tsikoyannis and Wei~\cite{tsikoyannis} considered a reactive one-dimensional system
with all the sites reactive in order to get more information about the
reactivity and selectivity in one-dimensional systems. They used a Markov
pure jump processes approach to model zeolitic diffusion and reaction as a
sequence of elementary jump events taking place in a finite periodic
lattice. Monte Carlo and approximate analytical solutions to the derived
Master Equation were developed to examine the effect of intracrystalline
occupancy on the macroscopic diffusional behavior of the system.
 One conclusion was that better results using analytical approach can be
obtained compared to DMC simulation results by including more
correlations between neighboring sites in regions of the systems
with high occupancy gradients and less correlations in regions
with low and no occupancy gradients. Starting from
Wei~\cite{tsikoyannis} results about correlations in Single-File
Systems, Okino and Snurr~\cite{okino} used a deterministic model
where each site was assumed to have equal activity towards
reaction. Doublet approximation was found to overpredict the
occupancy of the sites and the increasing mobility raised the
concentration of reactants in the pore.

 Using DMC simulations we have observed that even for infinitely fast diffusion, we
still have Single-File effects in the system.
 Instead of focusing on diffusion at different occupancies of the system, we
therefore concentrate in this paper on the reactivity of the system, studying the
reactivity of the system for different sets of kinetic parameters, the length of
the pipe and the distribution of the reactive sites.
  We analyse the situations when MF gives good results and when MF results
deviate strongly from the DMC simulations.
  We investigate the effect of the various model assumptions
made about diffusion, adsorption/desorption, and reaction  on
the overall behavior of the system. We look at the total loading, loading
with different components, generation of reaction products and occupancies
of individual sites as a function of the various parameters of a Single-File
System.

 In section \ref{sec:lev0} we specify our mathematical model for diffusion and reaction
in zeolites together with the theoretical background for the analytical and
simulation results. In section \ref {sec:lev1} we present the various results for
the simplified model without conversion. In section \ref{sec:lev2} we use
MF theory to solve the Master Equation governing the system behavior
for the case when all the sites have the same activity towards conversion.
 Similarly the results obtained using DMC simulations are
presented in section \ref{sec:lev3} and are compared with MF results. We
pay special attention to the infinitely fast diffusion case and to the
influence of the length of the pipe on the overall behavior of the system.
 In section \ref{sec:lev4} we analyze again the MF and simulation results but for the
case when only some of the sites are reactive. The influence of the position
and number of reactive sites on the reactivity and site occupancy of the
system is outlined. The last section summarizes our main conclusions.

\section {\label{sec:lev0} Theory}

  In this section we will give the theoretical background for our analytical
and simulation results. First we will specify our model and then we will
show that the defined system obeys a Master Equation.~\cite{kampen} We will
simulate the system governed by this Master Equation using DMC simulations. The rate equations used for the derivation of the
 analytical results are outlined.

\subsection {The Model}

    Because we are interested in reaction of molecules in
Single-File Systems, we call the system we are modelling, Single-File System with conversion.
    We model a Single-File System by a one-dimensional array of
sites, each possibly occupied by a single adsorbate. The sites are
numbered $\textsl {1, 2, \ldots, S}$. An adsorbate can only move if an
adjacent site is unoccupied. The sites could be reactive or unreactive and we note with $N_{reac}$
the number of reactive sites. A reactive site is the only place
where a reaction may take place.

\begin{figure*}
\centering
\subfigure {\epsfig {figure=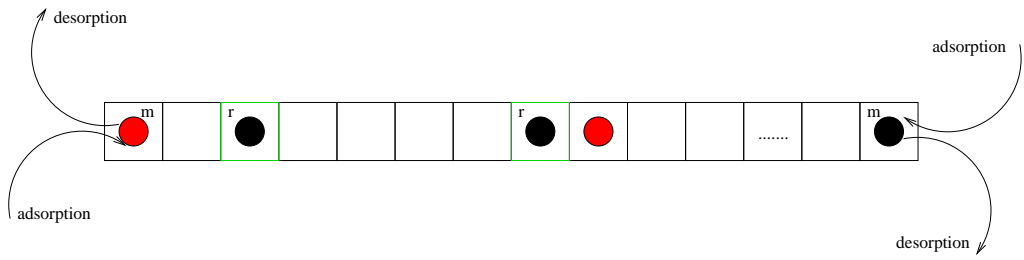, width=13.5cm} }
\caption {Picture of a Single-File System with two types of adsorbates,
$\rm A$(lighter colored) and $\rm B$(darker colored). The marginal sites are labeled with
$m$, and the reactive sites(lighter colored) with $r$. Adsoption of $\rm A$ and desorption
of $\rm A$ and $\rm B$ can take place only at the two marginal sites. An
$\rm A$ can transform into a $\rm B$ only on $r$ labeled sites.}
\label {singfile}
\end {figure*}
    We consider two types of adsorbates, $\rm A$ and $\rm B$, in our
model and we denote with $\rm X$ the site occupation of a site,
$\rm X$=($\rm *$, $\rm A$, $\rm B$), which stands for an empty
site, a site occupied by $\rm A$, or a site occupied by a $\rm B$,
respectively. The sites at the ends of the system are labeled with $m$, and
the reactive sites are labeled with $r$ (see figure~\ref{singfile}).
   We restrict ourselves to the following mono and bi-molecular transitions.
\\
\\
a) Adsorption and desorption
\\
\\
    Adsorption and desorption take place only at the two marginal sites
i.e., the left and rightmost sites at the ends of the system.
\\
\begin{center}
    ${\rm A}({\rm gas})$ + $\rm *_{\it m}$ $\longrightarrow$ ${\rm A}_{\it m}$
\\
    ${\rm A}_{\it m}$   $\longrightarrow$   ${\rm A}({\rm gas})$ + ${\rm *}_{\it m}$
\\
    ${\rm B}_{\it m}$ $\longrightarrow$ ${\rm B}({\rm gas})$ + $*_{\it m},$
\\
\end{center}
where $\it m$ denotes a marginal site. Note that there is no $\rm B$ adsorption.
$\rm B$'s are formed only by a reaction.
\\
\\
b) Diffusion
\\
\\
    In the pipe, particles are allowed to diffuse via hopping to
    vacant nearest neighbor sites.
\\
\begin{center}
    ${\rm A}_{\it n}$ + ${\rm *}_{\it n+1}$ $\longleftrightarrow$ ${\rm *}_{\it n}$ + ${\rm A}_{\it n+1}$
\\
    ${\rm B}_{\it n}$ + ${\rm *}_{\it n+1}$ $\longleftrightarrow$ ${\rm *}_{\it n}$ + ${\rm B}_{\it n+1},$
\\
\end{center}
where the subscripts are site indices: $n$=$\textsl{1, 2, \ldots, S-1}$.
\\
\\
c) Reaction
\\
\\
   An $\rm A$ can transform into a $\rm B$ at a reactive site.
\\
\begin {center}
    $\rm A_{\it r}$ $\longrightarrow$ $\rm B_{\it r}$.
\end{center}
       The initial state of the system is all that all sites are empty (no particles in
the pipe). In this paper we will only look at steady-state properties and
not to the time dependence of the system properties starting with no
particles.

\subsection { Master Equation}

  Reaction kinetics is described by a stochastic process. Every reaction has a microscopic
rate constant associated with it that is the probability per unit time that
the reaction occurs.
  Stochastic models of physical systems can be modelled by a Master
Equation.~\cite{kampen}

 By $\alpha$, $\beta$, we will indicate a particular configuration of the
system i.e., a particular way to distribute adsorbates over all the sites.
 $P_\alpha(t)$ will indicate the probability of finding the system in
configuration $\alpha$ at time $t$ and $W_{\alpha\beta}$ is the rate
constant of the reaction changing configuration $\beta$ to configuration
$\alpha$.

  The probability of the system being in configuration $\alpha$ at time
$t+dt$ can be expressed  as the sum of two terms. The first term is the
probability to find the system already in configuration $\alpha$ at time $t$
multiplied by the probability to stay in this configuration during $dt$.
 The second term is the probability to find the system in some other
configuration $\beta$ at time $t$ multiplied by the probability to go from
$\beta$ to $\alpha$ during $dt$.

\begin{equation}
P_{\alpha}(t+dt)=(1-dt\sum_{\beta} W_{\beta\alpha})P_{\alpha}(t) +
         dt\sum_{\beta}W_{\alpha\beta}P_{\beta}(t)
\end {equation}

By taking the limit $dt \to 0$ this equation reduces to a Master Equation:

\begin{equation}
  {dP_\alpha(t)\over dt}
  =\sum_{\beta}
   \left[W_{\alpha\beta}P_{\beta}(t)-W_{\beta\alpha}P_\alpha(t)\right].
\end{equation}

Analytical results can be derived as follow. The value of a property $X$ is a weighted average over the values
$X_{\alpha}$ which is the value of $X$ in configuration $\alpha$:

\begin{equation}
   \langle X \rangle=\sum_{\alpha}P_{\alpha}X_{\alpha}.
\end{equation}

  From this follows the rate equation
\begin{equation}
\begin{split}
 {{d\langle X \rangle}\over{dt}}
  & =\sum_{\alpha}{dP_{\alpha}\over{dt}}X_{\alpha}\cr
  &=\sum_{\alpha\beta}[W_{\alpha\beta}P_{\beta}-W_{\beta\alpha}P_{\alpha}]X_{\alpha}\cr
  &=\sum_{\alpha\beta}W_{\alpha\beta}P_{\beta}(X_{\alpha}-X_{\beta}).
\end{split}
\end{equation}

\subsection {Dynamic Monte Carlo}

  Because it might be not always possible to solve the Master Equation
analytically, DMC methods allow us to simulate the system governed by the
Master Equation over time.
  We simplify the notation of the Master Equation by defining a matrix $\bf W$
containing the rate constants $W_{\alpha\beta}$, and a diagonal matrix $\bf R$ by
${R}_{{\alpha}{\beta}}\equiv \sum_{\gamma}W_{\gamma\beta}$, if ${\alpha}={\beta}$,
and 0 otherwise.
 If we put the probabilities of the configurations $P_{\alpha}$ in a vector
$\bf P$, we can write the Master Equation as

\begin {equation}
 {d{\bf P}\over{dt}}=-({\bf R}-{\bf W}){\bf P}.
\end{equation}
where $\bf R$ and $\bf W$ are assumed to be time independent.
We also introduce a new matrix $\bf Q$, ${\bf Q}(t) \equiv \exp[-{\bf R}t].$

This matrix is time dependent by definition and we can rewrite the Master
Equation in the integral form

\begin{equation}
{\bf P}(t)={\bf Q}(t){\bf P}(0)+\int_0^tdt^{\prime}{\bf Q}(t-t^{\prime}){\bf W}{\bf P}(t^{\prime}).
\end{equation}
By substitution we get of the right-hand-side  for $P(t^{\prime})$

\begin{equation}
\begin{split}
{\bf P}(t)
  & =[{\bf Q}(t)\cr
  & +
     \int_0^t dt^{\prime}{\bf Q}(t-t^{\prime}){\bf W}{\bf Q}(t^{\prime})\cr
  & +
    \int_0^tdt^{\prime}\int_0^{t^{\prime}}dt^{\prime\prime}{\bf Q}(t-t^{\prime}){\bf W}{\bf Q}(t^{\prime}-t^{\prime\prime})
    {\bf W}{\bf Q}(t^{\prime\prime})\cr
  & +\ldots]{\bf P}(0).
\end{split}
\end{equation}

 Suppose at $t=0$ the system is in configuration $\alpha$ with probability
$P_{\alpha}(0)$. The probability that, at time $t$, the system is still in
configuration $\alpha$ is given by
$Q_{\alpha\alpha}(t)P_{\alpha}(0)=\exp(-R_{\alpha\alpha}t)P_{\alpha}(0)$.
 This shows that the first term represents the contribution to the
probabilities when no reaction takes place up to time $t$. The matrix $\bf W$
determines how the probabilities change when a reaction takes place. The
second term represents the contribution  to the probabilities when no
reaction takes place between times $0$ and $t^{\prime}$, some reaction takes
place at time $t^{\prime}$, and then no reaction takes place between
$t^{\prime}$ and $t$. The subsequent terms represent contributions when two,
three, four, etc. reactions take place.
 The idea of the DMC method is not to compute probabilities
$P_{\alpha}(t)$ explicitly, but to start with some particular configuration,
representative for the initial state of the experiment one wants to
simulate, and then generate a sequence of other configurations with the
correct probability.
 The method generates a time $t^{\prime}$ when the first reaction occurs
according to the probability distribution $1-\exp[-R_{\alpha\alpha}t]$.
At time $t^{\prime}$ a reaction takes place such that a new configuration
${\alpha}^{\prime}$ is generated by picking it out of all possible new
configurations $\beta$ with a probability proportional to
$W_{{\alpha}^{\prime}{\alpha}}$. At this point we can proceed by repeating
the previous steps, drawing again a time for a new reaction and a new
configuration.

\section { Results and Discussion }

\subsection {\label{sec:lev1} No conversion}

  We mention in this section various results for the system without
conversion. These results can be derived analytically. The derivations are
not difficult, so for completeness we give them in the appendix. We will use
the results when we deal with the system with conversion.

   In a Single-File System without conversion, the relevant processes to
describe are adsorption, desorption and diffusion.
   So, $ W_{\alpha\beta} $ is given by
\begin{equation}
  W_{\alpha\beta}
  =W_{\rm ads}\Delta_{\alpha\beta}^{({\rm ads})}
  +W_{\rm des}\Delta_{\alpha\beta}^{({\rm des})}
  +W_{\rm diff}\Delta_{\alpha\beta}^{({\rm diff})},
\end{equation} where $\Delta_{\alpha\beta}^{({\rm rx})}$  equals 1 if a
reaction of type ``rx'' can transform the system from $\beta$ to $\alpha$,
and equals 0 otherwise. $W_{\rm ads}$, $W_{\rm des}$, $W_{\rm diff}$ are the
rate constants of adsorption, desorption and diffusion respectively.

 If we substitute  expression (8) into the Master Equation (2), we get
\begin{align}
  {dP_\alpha\over dt}
  &=W_{\rm ads}\sum_\beta
   \left[\Delta_{\alpha\beta}^{({\rm ads})}P_\beta
        -\Delta_{\beta\alpha}^{({\rm ads})}P_\alpha\right] \nonumber \\
   &+W_{\rm des}\sum_\beta
    \left[\Delta_{\alpha\beta}^{({\rm des})}P_\beta
    -\Delta_{\beta\alpha}^{({\rm des})}P_\alpha\right]  \\
   &+W_{\rm diff}\sum_\beta
     \left[\Delta_{\alpha\beta}^{({\rm diff})}P_\beta
     -\Delta_{\beta\alpha}^{({\rm diff})}P_\alpha\right]. \nonumber
\end{align}

 Using this expression we can show that
  {\sl when the system is in steady state then the probability of
    finding the system in a certain configuration depends only on the
    number of particles in the system.}
\begin{equation}
  P_\alpha=q\big(N_{\alpha}\big)
\end{equation}
where $N_{\alpha}$ is the number of particles in configuration $\alpha$.

The expression for $q(N)$ is:
\begin{equation}
  q(N)=\left[{W_{\rm des}\over W_{\rm des}+W_{\rm ads}}\right]^S
       \left[{W_{\rm ads}\over W_{\rm des}}\right]^N.
\end{equation}
Note that diffusion has here no effect on steady-state properties.

The loading of the pipe, defined as the average number of particles per
site, is then
\begin{equation}
  {Q_{\rm A}={1\over{S}}{{\sum_{N=0}^SN\,p(N)}}}={W_{\rm ads}\over W_{\rm ads}+W_{\rm des}},
\end{equation}
where $p(N)$ is the probability that there are $N$ particles in the system.
Note again that diffusion doesn't influence the steady-state loading.

The standard deviation, i.e., the fluctuation in the number of
particles is then:
\begin{equation}
\begin{split}
 \sqrt{{\sigma}^2}
 & =\sqrt{\sum_{N=0}^SN^2\,p(N)-\left[\sum_{N=0}^SN\,p(N)\right]^2}\cr
 &=\sqrt{{W_{\rm ads}W_{\rm des}\over(W_{\rm des}+W_{\rm ads})^2}S}.
\end{split}
\end{equation}

  To determine how the parameters of the system influence the
kinetics of the system, we are interested in the correlation in the occupancy
between neighboring sites. We look at one site occupancy and at two
sites occupancies.
  We  denote by $\langle {\rm A}_n \rangle$ the probability that an $\rm A$ is at site $n$
and with $\langle {\rm A}_n {\rm A}_{n+1} \rangle$ the probability to have an
$\rm A$ at site $n$ and one at site $n+1$.

   One and two-site probabilities can be derived from the fact that all
configurations with the same number of particles have equal probability and
the expressions for $q(N)$. We find
\begin{equation}
  {\langle {\rm A}_n \rangle}={W_{\rm ads}\over{W_{\rm ads} + W_{\rm des}}},
\end{equation}
and
\begin{equation}
 \langle {\rm A}_n {\rm A}_{n+1} \rangle
  = {\left[{W_{\rm ads}\over{W_{\rm ads}+W_{\rm des}}}\right]} ^2,
\end{equation}

 Note that this probability does not depend on the site, all sites have equal
probability to be occupied and that there is no correlation between the
occupation of neighboring sites.
 Again diffusion doesn't influence these properties. Note also that these
expressions are the same as for a model in which particles are allowed to pass each
other.

\subsection {\label{sec:lev2} All sites reactive}

  We look first at the situation with all sites reactive: i.e., conversion of an
$\rm A$ into a $\rm B$ particle can take place at any site including the marginal sites.
   For simplicity we consider $W_{\rm desA}$=$W_{\rm desB}$=$W_{\rm des}$, and also
$W_{\rm diffA}$=$W_{\rm diffB}$=$W_{\rm diff}$.
  We will be looking at the total loading ($Q$), the total loading of $\rm A$'s ($Q_{\rm A}$),
the total loading of $\rm B$'s ($Q_{\rm B}$), the number of $\rm B$'s produced per unit time
(${B}_{prod}$), and how the distribution of $\rm A$'s and $\rm B$'s varies from site to site
($\langle {\rm A}_n \rangle$ and $\langle {\rm B}_n \rangle$).

  Note that the total loading of the pipe for the model with conversion is
the same as for the model without conversion
\begin{equation}
Q={W_{\rm ads}\over{W_{\rm ads}+W_{\rm des}}}.
\end{equation}
The loadings and the production of $\rm B$'s can easily be derived  from the
probabilities $\langle {\rm A}_n \rangle$ and $\langle {\rm B}_n \rangle$ so we first focus on them. For a non-marginal
site we can write
\begin{equation}
  {d \langle {\rm A}_n \rangle \over{dt}}={R_n}^{({\rm A,diff})}+R_n^{(\rm rx)},
\end{equation}
where ${R_n}^{\rm (A,diff)}$ is the rate of diffusion of $\rm A$ from and to site
$n$, and ${R_n}^{\rm (rx)}$ is the rate of conversion of $\rm A$ to $\rm B$ on site $n$. The conversion
takes place at one site and is therefore easier to handle than the
diffusion. Using equation (4) we have
\begin{equation}
{R_n}^{\rm (rx)}=W_{\rm rx}\sum_{\alpha\beta}{{\Delta}_{\alpha\beta}}^{\rm (rx)}P_{\beta}({\rm A}_{n\alpha}-{\rm A}_{n\beta}),
\end{equation}
where ${\rm A}_{n\alpha}=1$ if site $n$ is occupied by an $\rm A$ in configuration $\alpha$
and ${\rm A}_{n\alpha}=0$ if not.
 We have ${\rm A}_{n\alpha } - {\rm A}_{n\beta}\not=0$ if there is an $\rm A$ at site $n$ in
configuration $\beta$ (${\rm A}_{n\beta}$=1) that has reacted to a $\rm B$ leading to
configuration $\alpha$ (${{\rm A}_{n\alpha}=0}$). This gives us
\begin{equation}
{R_n}^{\rm (rx)}=-W_{\rm rx}{{\sum_{\beta}^{}}^{\prime}P_{\beta}}=-W_{\rm rx}\langle {\rm A}_n \rangle,
\end{equation}
where the prime restricts the summation to those $\beta$'s with
${\rm A}_{n\beta}=1$. For the diffusion we similarly get
\begin{equation}
{R_n}^{(\rm A,diff)}=W_{\rm diff}\sum_{\beta}{{\Delta}_{\alpha\beta}}^{(\rm A,diff)}P_{\beta}({\rm A}_{n\alpha}-{\rm A}_{n\beta}).
\end{equation}
There are four ways in which ${\rm A}_{n\alpha}-{\rm A}_{n\beta}\not=0$ and
${\Delta_{\alpha\beta}}^{\rm (A,diff)}\not=0$ in $\beta$: there is an $\rm A$ at site
$n$ that can move to site $(n-1)$, there is an $\rm A$ at site $n$ that can move to
$(n+1)$, there is an $\rm A$ at site $(n-1)$ that can move to site $n$ and there is an
$\rm A$ at site $(n+1)$ that can move to site $n$. In all cases we have
${{\Delta}_{\alpha\beta}}^{\rm (A,diff)}=1.$ In the first two cases we have
${\rm A}_{n\alpha}-{\rm A}_{n\beta}=-1$ and in the last two we have
${\rm A}_{n\alpha}-{\rm A}_{n\beta}=1.$ The summation over $\beta$ in the first case is
restricted to configurations with an $\rm A$ at site $n$ and a vacant site
$(n-1)$.
 This gives a term $-W_{\rm diff}\langle *_{n-1}{\rm A}_n \rangle.$ The other cases give terms
$-W_{\rm diff} \langle {\rm A}_n *_{n+1} \rangle$, $W_{\rm diff} \langle {\rm A}_{n-1} *_n \rangle $, and
$W_{\rm diff} \langle *_n {\rm A}_{n+1} \rangle.$
 The rate equations then becomes
\begin{equation}
\begin{split}
 { d\langle {\rm A}_n \rangle\over{dt}}
&=W_{\rm diff}[-\langle {\rm A}_n *_{n-1}\rangle-\langle *_{n-1} {\rm A}_n\rangle+\langle {\rm A}_{n-1}*_n\rangle \cr
&+\langle *_n {\rm A}_{n+1}\rangle] - W_{\rm rx}\langle {\rm A}_n\rangle.
\end{split}
\end{equation}
 For $\langle {\rm B}_n\rangle$ we get similarly
\begin{equation}
\begin{split}
 { d\langle {\rm B}_n\rangle\over{dt}}
&=W_{\rm diff}[-\langle {\rm B}_n *_{n-1}\rangle-\langle *_{n-1} {\rm B}_n\rangle+\langle {\rm B}_{n-1}*_n\rangle \cr
&+\langle *_n {\rm B}_{n+1}\rangle] + W_{\rm rx}\langle {\rm A}_n\rangle.
\end{split}
\end{equation}
 The marginal sites have  also adsorption and desorption. They can be dealt
with as the conversion. The rate equations for $\rm A$ are
\begin{equation}
\begin{split}
{d\langle {\rm A}_1\rangle\over{dt}}
&=W_{\rm diff}[-\langle {\rm A}_1 *_{2}\rangle+\langle *_1
 {\rm A}_{2}\rangle]+W_{\rm ads}\langle *_1\rangle \cr
&-W_{\rm des}\langle{\rm A}_1\rangle - W_{\rm rx}\langle {\rm A}_1\rangle,\cr
{d\langle {\rm A}_S\rangle\over{dt}}
& =W_{\rm diff}[-\langle {\rm A}_S *_{S-1}\rangle+\langle *_S
 {\rm A}_{S-1}\rangle]+W_{\rm ads}\langle *_S\rangle \cr
&-W_{\rm des}\langle {\rm A}_S\rangle - W_{\rm rx}\langle
{\rm A}_S\rangle,\cr
\end{split}
\end{equation}
and the rate equations for $\rm B$
\begin{equation}
\begin{split}
 {d\langle{\rm B}_1\rangle\over{dt}}
&=W_{\rm diff}[-\langle {\rm B}_1 *_{2}\rangle+\langle *_1 {\rm B}_{2}\rangle]-W_{\rm des}\langle {\rm B}_1\rangle+W_{\rm rx}\langle{\rm A}_1\rangle,\cr
{d\langle {\rm B}_S\rangle\over{dt}}
&=W_{\rm diff}[-\langle {\rm A}_S *_{S-1}\rangle+\langle *_S {\rm A}_{S-1}\rangle]-W_{\rm des}\langle {\rm B}_1\rangle\cr
&+W_{\rm rx}\langle
{\rm A}_S\rangle.
\end{split}
\end{equation}
Note that these coupled sets of differential equations are exact.

\subsubsection { Mean Field results }

   We will now look at the loadings $Q_{\rm A}$ and $Q_{\rm B}$ and the site occupation
probabilities $\langle {\rm A}_n \rangle$  and $\langle {\rm B}_n \rangle$. We will first determine steady-state properties using the (MF)
approximation: i.e, we put $\langle {\rm A}_n *_{n+1} \rangle$=$\langle {\rm A}_n  \rangle
\langle *_{n+1} \rangle$ etc. in the rate equations. This gives us

\begin{equation}
\begin{split}
  0&={W_{\rm diff}W_{\rm des}\over W_{\rm ads}+W_{\rm des}}\left[
     \expec{{\rm A}_{n+1}}+\expec{{\rm A}_{n-1}}
    -2\expec{{\rm A}_n}\right]
   -W_{\rm rx}\expec{{\rm A}_n},\cr
  0&={W_{\rm diff}W_{\rm des}\over W_{\rm ads}+W_{\rm des}}\left[
     \expec{{\rm B}_{n+1}}+\expec{{\rm B}_{n-1}}
    -2\expec{{\rm B}_n}\right]
   +W_{\rm rx}\expec{{\rm A}_n},\cr
  0&={W_{\rm diff}W_{\rm des}\over W_{\rm ads}+W_{\rm des}}\left[
    \expec{{\rm A}_2}-\expec{{\rm A}_1}\right]
   -W_{\rm rx}\expec{{\rm A}_1}
   -W_{\rm des}\expec{{\rm A}_1} \cr
   & +{W_{\rm ads}W_{\rm des}\over W_{\rm ads}+W_{\rm des}},\cr
  0&={W_{\rm diff}W_{\rm des}\over W_{\rm ads}+W_{\rm des}}\left[
    \expec{{\rm B}_2}-\expec{{\rm B}_1}\right]
   +W_{\rm rx}\expec{{\rm A}_1}
   -W_{\rm des}\expec{{\rm B}_1},\cr
  0&={W_{\rm diff}W_{\rm des}\over W_{\rm ads}+W_{\rm des}}\left[
    \expec{{\rm A}_{S-1}}-\expec{{\rm A}_S}\right]
   -W_{\rm rx}\expec{{\rm A}_S}
   -W_{\rm des}\expec{{\rm A}_S} \cr
   & +{W_{\rm ads}W_{\rm des}\over W_{\rm ads}+W_{\rm des}},\cr
  0&={W_{\rm diff}W_{\rm des}\over W_{\rm ads}+W_{\rm des}}\left[
    \expec{{\rm B}_{S-1}}-\expec{{\rm B}_S}\right]
   +W_{\rm rx}\expec{{\rm A}_S}
   -W_{\rm des}\expec{{\rm B}_S}.\cr
\end{split}
\end{equation}
We have used here the probability for a site to be vacant that we have
determined for the case without conversion.

 We note that these equations are identical to the MF equations of a system
in which the particles can move independently with a rate constant for diffusion equal
to ${W_{\rm diff}W_{\rm des}}/{(W_{\rm des}+W_{\rm ads})}$.
 This means that the MF does not  really model the non-passing that
characterizes a Single-File System.

 The continuum limit of the MF equation
is
\begin{equation}
 \begin{pmatrix}
    {\partial a}/{\partial t}\cr
    {\partial b}/{\partial t}\cr
 \end{pmatrix}
 =D\begin{pmatrix}
   1-b & a \cr
   b& 1-a \cr
   \end{pmatrix}
   \begin{pmatrix}
    {\partial^2 a}/ \partial x^2\cr
    {\partial^2 b}/ \partial x^2\cr
   \end{pmatrix}
  + W_{\rm rx}\begin{pmatrix}
   -a \cr
    a \cr
    \end{pmatrix},
\end{equation}
where $a$=$a(x,t)$ is the probability distribution of $\rm A$'s (a
similar definition holds for $b$), and $D$=${{W}_{\rm diff}}d^2$,
with $d$ the distance between neighboring sites(see appendix).
 These are the equations that are normally used to describe diffusion in
Single-File Systems.~\cite{coppens2,krishna,paschek1,paschek2} As
this equation is derived from the MF equations, it has the same
drawback; i.e., the Single-File behavior is only incorporated by
the reduction of the diffusion, but it does effectively allow for
passing of particles. This shows up as so-called counter diffusion
of A's and B's.~\cite{krishna,paschek1,paschek2}

  We see that equations (25) are linear and we can solve them at least numerically. We think however that
it is worthwhile to use an analytical approach. We consider the
{\it ansatz\/}
\begin{equation}
  \expec{{\rm A}_n}\propto x^n
\end{equation}
in the steady-state equations (25) for $\langle {\rm A}_n\rangle.$
This leads to
\begin{equation}
  x^2-2(1+\alpha)x+1=0,
\end{equation}
with
\begin{equation}
  \alpha={{W_{\rm rx}\over 2W_{\rm diff}}{W_{\rm des}+W_{\rm ads}\over{W_{\rm des}} }}.
\end{equation}
The quadratic equation yields two solutions $x_1$ and $x_2$ with
${x_2}={x_1^{-1}}.$ We have $x_1=x_2=1$ only when $\alpha=0$, i.e. when
$W_{\rm rx}=0.$ We will therefore assume $\alpha>0$ and $x_1<1.$ Then
\begin{equation}
 x_{1}=(1+\alpha)-\sqrt{\alpha(\alpha+2)}.
\end{equation}
We can write then the solution $\langle {\rm A}_n\rangle=a_1(x_1)^n + a_2(x_1)^{S+1-n}$. The
symmetry in the occupancy of the pipe $\langle {\rm A}_n\rangle=\langle
{\rm A}_{S+1-n}\rangle$ yields $a_1=a_2=a$. So, the general solution for the steady state
has the form:
\begin{equation}
  \expec{{\rm A}_n}=a(x_1^n+x_1^{S+1-n}).
\end{equation}

\begin {figure}
\centering
\subfigure {\epsfig {figure=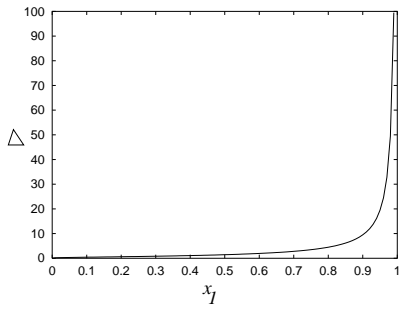, width=5.5cm} }
\caption {
 The characteristic length $\Delta$ as a function of $x_1$.
         }
\label{Fign2}
\end {figure}

 The coefficient $a$ is to be determined from the equations for the marginal
sites in the set of equations(25).
 In the left side of the system $n$ is small and $(S+1-n)$ is large. Because
$x_1<1$ we can neglect the second term in equation (21) and
$\expec{{\rm A}_n}\propto {x_1}^n.$ This means that the
probability of finding an $\rm A$ at site in the left-hand-side of
the system is an exponentially decreasing function of the site
index. If we write $\langle {\rm A}_n \rangle \propto
e^{-{n\over{\Delta}}}$, we find $\Delta=-{1/{ln(x_1)}}$ for the
characteristic length of the decrease. The logarithm makes this
length only a slowly varying function of the rate constants(see
figure ~\ref{Fign2}). When $W_{\rm diff}$ becomes larger, $\alpha$
approaches 0, $x_1$ approaches 1 and $\Delta$ diverges. Note that
this is a MF result. We will see that in the simulations $\Delta$
remains finite. Also when the conversion is slow more $\rm A$'s
are found away from the marginal sites. The second factor in the
expression for $\alpha$ equals the reciprocal of a site being
vacant. Low loading leads to a smaller $\alpha$ than high loading.
Because of the vacancies the $\rm A$'s can penetrate farther into
the system before being converted. For slow conversion or fast
diffusion $\alpha$ is small and $\Delta$ can be approximated by
\begin{equation}
\Delta=\sqrt{ {W_{\rm diff}\over{W_{\rm rx}} } {W_{\rm des}\over{W_{\rm des}+W_{\rm ads}}} }.
\end{equation}


\vspace*{-0cm}

\vspace*{-0cm}
\begin {figure}
\centering
\subfigure {\epsfig {figure=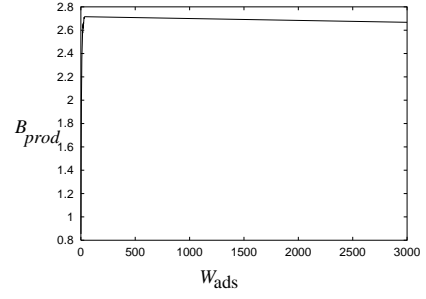, width=5.5cm} }
\caption {
 ${B}_{prod}$ per unit time at one marginal site as a
function of $W_{\rm ads}$ for $S$=${N_{reac}}$=30, $W_{\rm
des}$=0.8, $W_{\rm diff}=2$ and $W_{\rm rx}=0.4$.
           }
\label{Fign3}
\end {figure}

 The total loading with $\rm A$'s, $Q_{\rm A}$, is

\begin{equation}
 Q_{\rm A}={1\over{S}}\sum_{n=1}^S \langle {\rm A}_n\rangle
\end{equation}
so, the expression for $Q_{\rm A}$ is

\begin{equation}
Q_{\rm A}={a\over{S}}\sum_{n=1}^S[{x_1}^n + {x_1}^{S+1-n}]
    ={{2a}\over{S}} {{x_1{(1-{x_1}^S)}}\over{1-x_1}},
\end{equation}
\begin{equation}
Q_{\rm B}=Q-Q_{\rm A}.
\end{equation}
The total production of $\rm B$'s is
\begin{equation}
{B}_{prod}=W_{\rm rx}Q_{\rm A}S=2aW_{\rm rx} {{x_1({1-{x_1}^S})}\over{1-x_1}}.
\end{equation}

\subsubsection {\label{sec:lev3} Simulation results }

  We present now the results for different sets of parameters and we compare
them with MF results. Because we can see from equation (36) that
larger pipes don't increase the productivity of the system, we
consider for the comparisons of the results a system size
$\textsl{S}=30$.
  We have considered separately the sets of parameters in Table~\ref{tab:Table1}.

\begin {table}
\begin {center}
\begin{tabular}{|l|cccc|}
\hline set & $W_{\rm ads}$ & $W_{\rm des}$ & $W_{\rm diff}$ & $W_{\rm rx}$\\
\hline  a)& 0.2 & 0.8 & 0.05 & 0.01\\
 b)& 0.2 & 0.8 & 0.05 & 0.1\\
 c)& 0.2 & 0.8 & 2 & 0.1\\
 d)& 0.2 & 0.8 & 1 & 2\\
 e)& 0.2 & 0.8 & 10 & 2\\
 f)& 0.8 & 0.2 & 0.05 & 0.01\\
 g)& 0.8 & 0.2 & 0.05 & 0.1\\
 h)& 0.8 & 0.2 & 2 & 0.1\\
 i)& 0.8 & 0.2 & 1 & 2\\
 j)& 0.8 & 0.2 & 10 & 2\\
\hline
\end{tabular}
\end{center}
\caption {\label{tab:Table1} The sets of parameters used for the simulations}
\end{table}

\begin {figure*}
\centering
\subfigure {\epsfig {figure=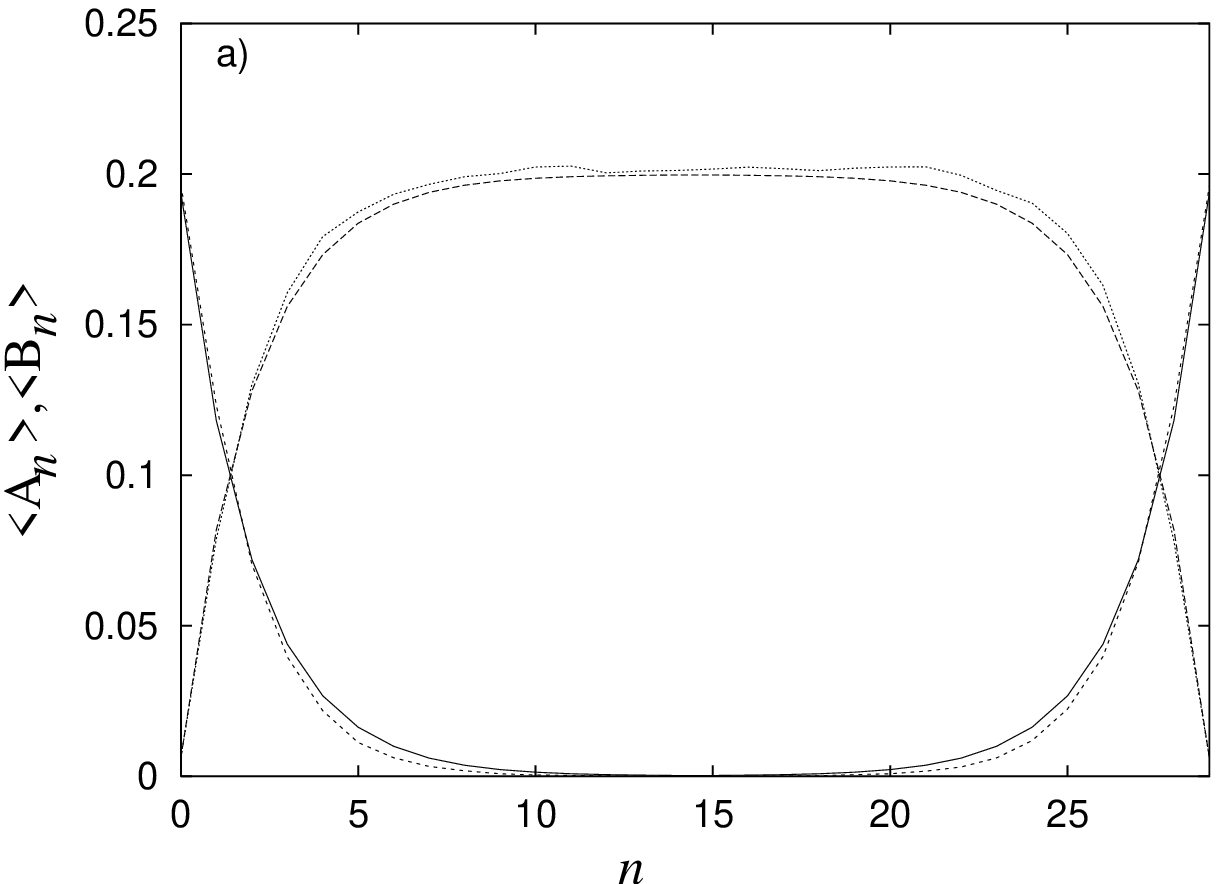, width=4.5cm} }
\subfigure {\epsfig {figure=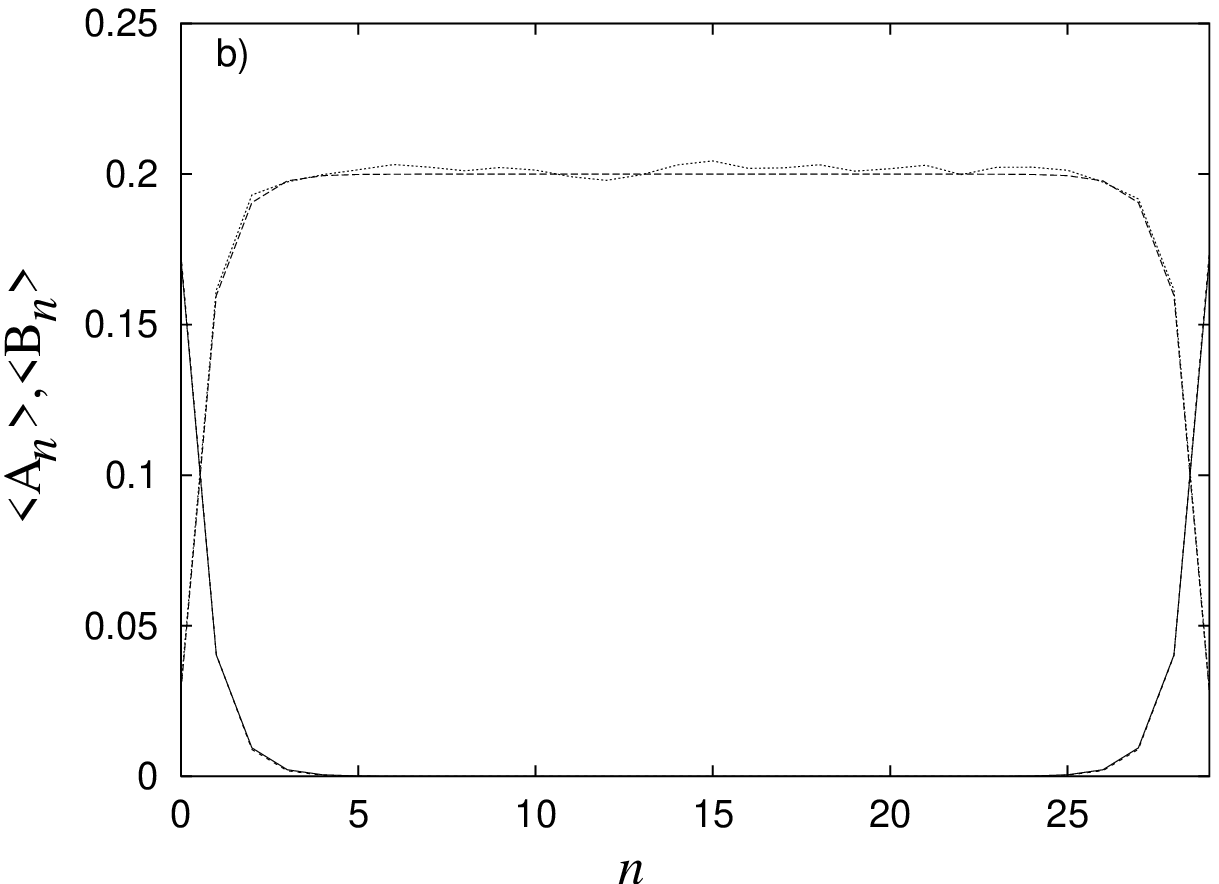, width=4.5cm} }
\subfigure {\epsfig {figure=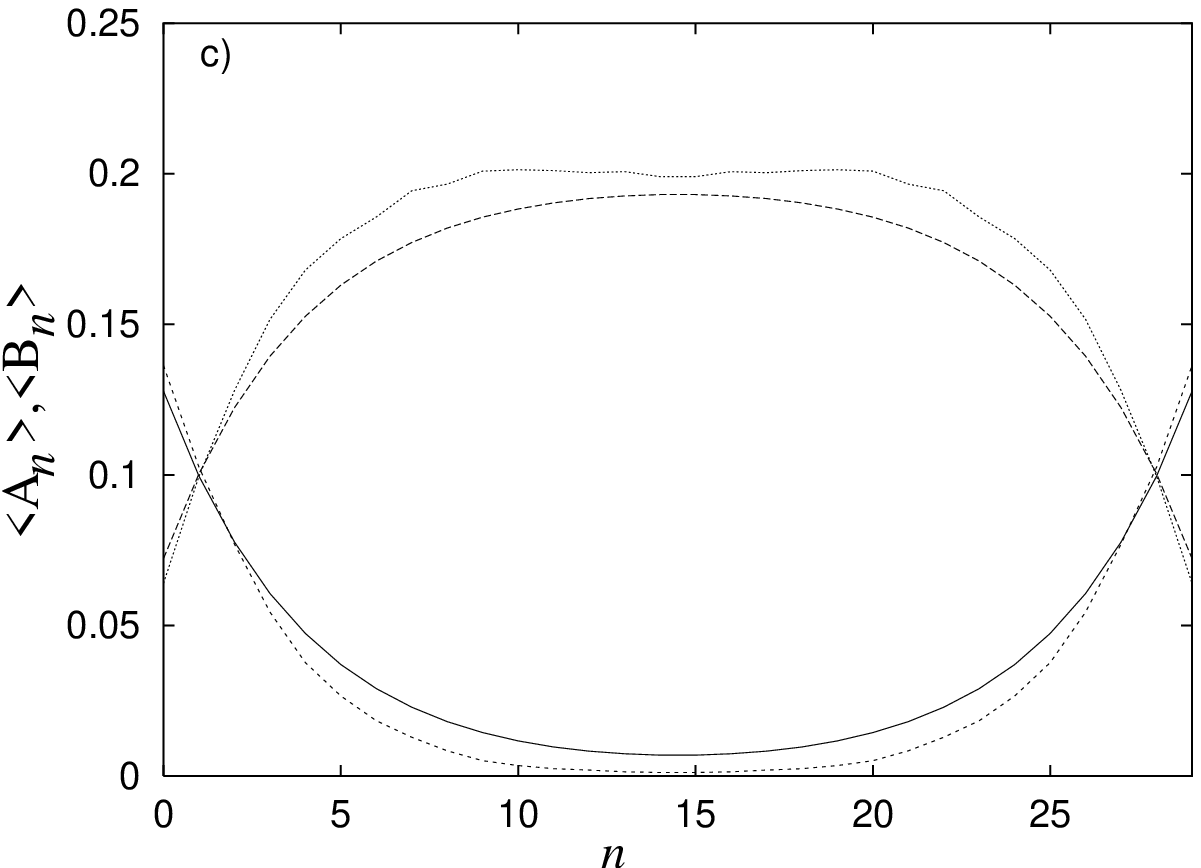, width=4.5cm} }
\subfigure {\epsfig {figure=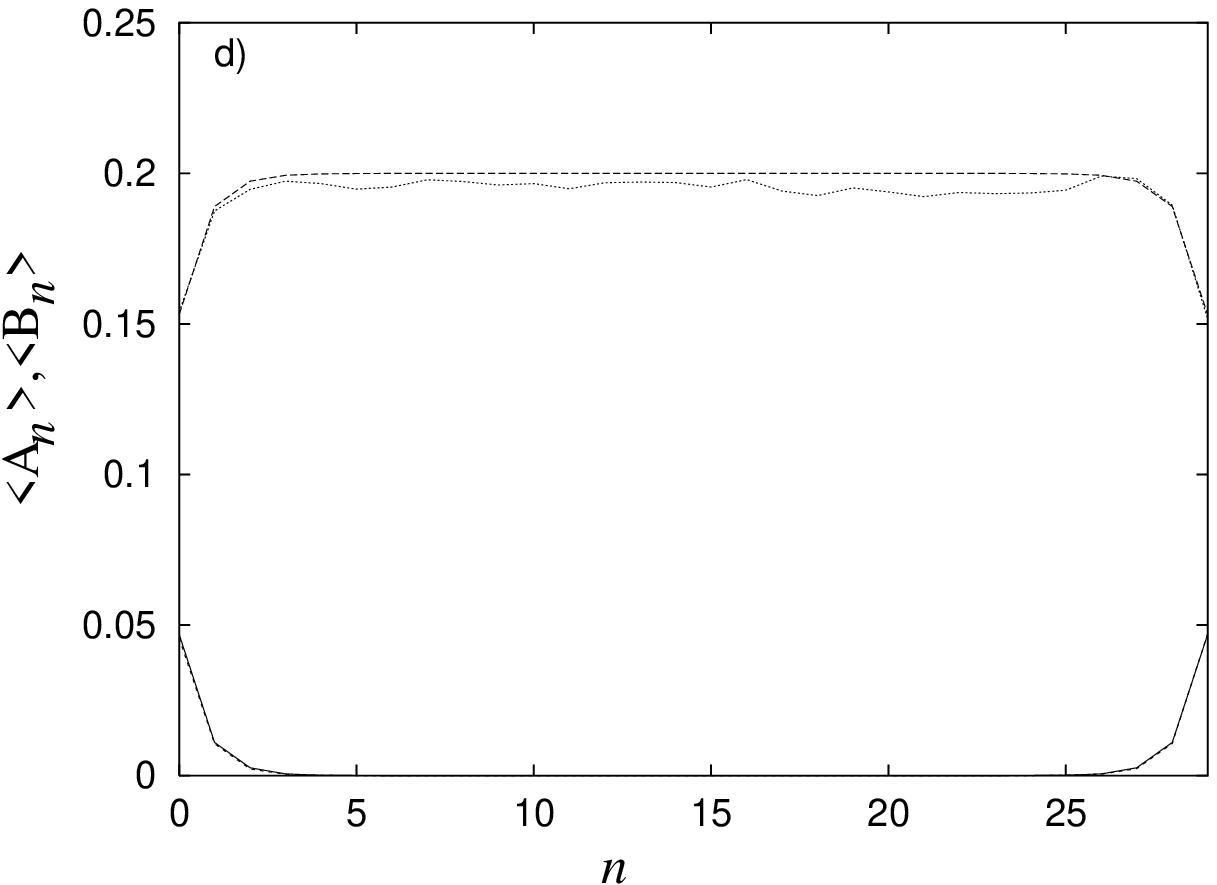, width=4.5cm} }
\subfigure {\epsfig {figure=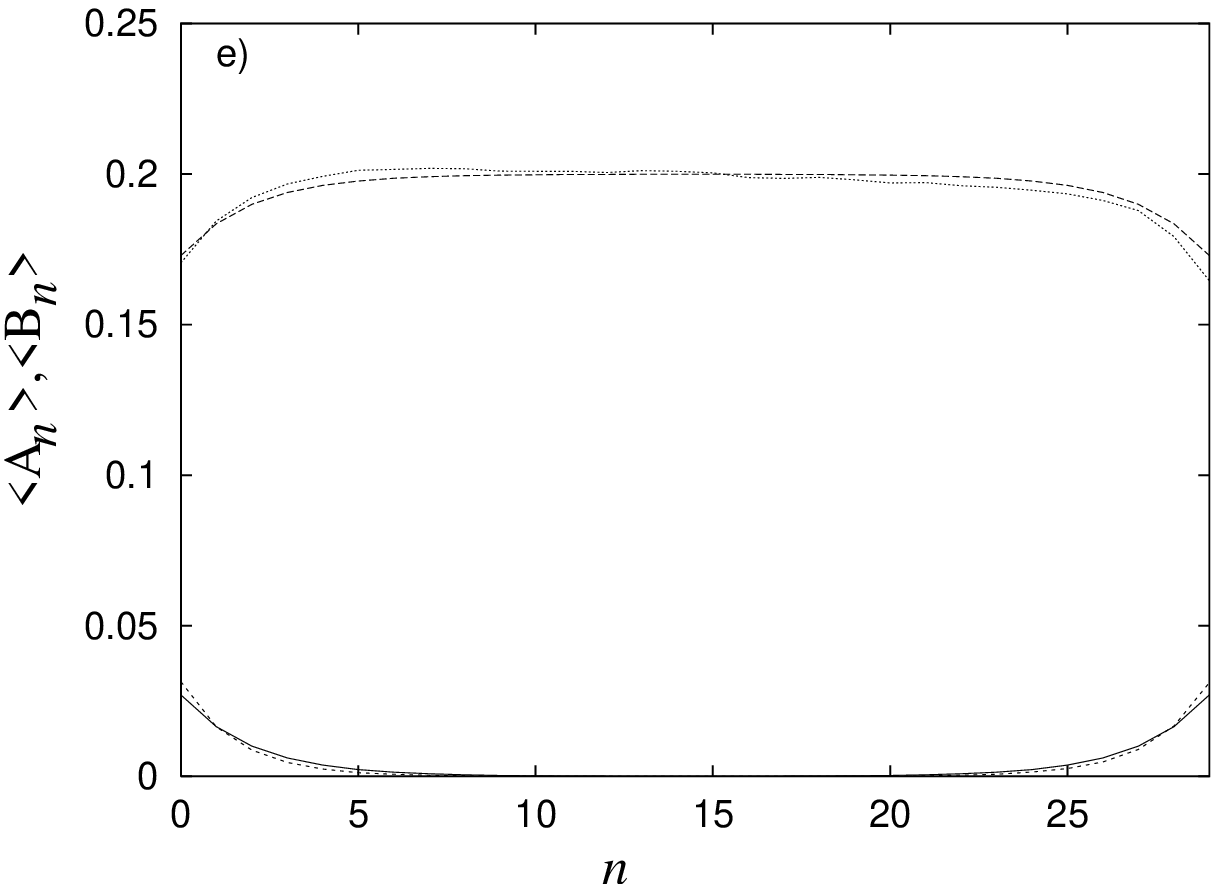, width=4.5cm} }
\caption
{
  The site occupancy with $\rm A$ ($\langle {\rm A}_n \rangle$) and $\rm B$
($\langle {\rm B}_n \rangle$) as a function on site number for
cases a, b, c, d, e when $S=N_{reac}=30$. The continuous line and
the corresponding symmetric line represent MF results for $\langle
{\rm A}_n \rangle$ and $\langle {\rm B}_n \rangle$ respectively.
The dashed lines represent DMC results for $\langle {\rm A}_n
\rangle$ and $\langle {\rm B}_n \rangle$. $\langle {\rm A}_n
\rangle$ is decreasing towards the middle of the pipe while
$\langle {\rm B}_n \rangle$ is increasing.
}
\label{Figurina}
\end {figure*}

\begin {figure*}
\centering
\subfigure {\epsfig {figure=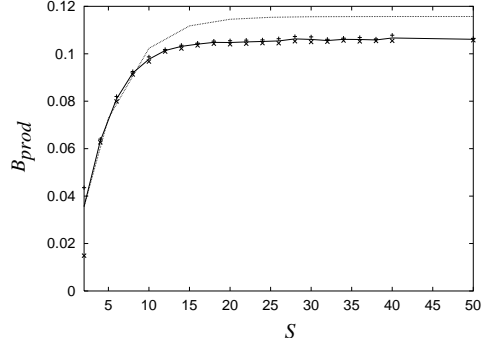, width=6.5cm} }
\caption { $\rm B$ production as a function on the length of the pipe for
           $W_{\rm ads}$=0.2, $W_{\rm des}$=0.8, $W_{\rm diff}$=2, and $W_{\rm rx}$=0.1.
           The marked line represents the DMC results and the dashed
           line represents the MF results. }
\label{Fig4}
\end {figure*}

 The sets of parameters from a) to e) are for the cases of low loading and from f) to j)
for the high loading. The parameters in the table describe the
following situations:
a) and f) for very slow reaction and slow diffusion;
b) and g) for slow reaction and slow diffusion;
c) and h) for slow reaction and fast diffusion;
d) and i) for fast reaction and slow diffusion;
e) and j) for fast reaction and fast diffusion.
\\
\begin{table}
\begin{center}
\begin{tabular}{|l| p{1.5cm} p{1.7cm} p{1.5cm} p{1.5cm} p{0.8cm}|}
\hline \multicolumn {1}{|c}{}  &\multicolumn {2}{c}{$Q_{\rm A}$} & \multicolumn {2}{c}{${B}_{prod}$}& \multicolumn {1}{c|} {Q}\\
\hline  set & MF & Sim & MF & Sim & Sim\\
\hline a)& 0.0330 & 0.0318 & 0.0099 & 0.0100 & 0.209\\
 b)& 0.0149 & 0.0148 & 0.0491 & 0.0472 & 0.198\\
 c)& 0.0385 & 0.0342 & 0.1156 & 0.1024 & 0.204\\
 d)& 0.0040 & 0.0041 & 0.2449 & 0.2463 & 0.200\\
 e)& 0.0046 & 0.0044 & 0.2767 & 0.2729 & 0.201\\
 f)& 0.0798 & 0.0748 & 0.0239 & 0.0235 & 0.795\\
 g)& 0.0376 & 0.0373 & 0.1129 & 0.1157 & 0.804\\
 h)& 0.0598 & 0.0486 & 0.1796 & 0.1406 & 0.802\\
 i)& 0.0048 & 0.0049 & 0.2931 & 0.2943 & 0.797\\
 j)& 0.0050 & 0.0049 & 0.3013 & 0.2957 & 0.801\\
\hline
\end{tabular}
\end{center}
\caption {\label{tab:Table2} Simulation and MF results for $Q_{\rm A}$ and ${B}_{prod}$ for all the sets of parameters}
\end{table}
  We can see from Table~\ref{tab:Table2} that the simulation and MF results match for all
the cases except the cases when we have low reaction rates and fast
diffusion for both low and high loading. In these cases MF overestimates the amount
of $\rm A$'s in the pipe, and consequently overestimates the B production.
In figure~\ref{Figurina} we have the site occupancy with $\rm A$ and $\rm B$
both from the simulations and MF.
 We again see that the MF and the simulation results agree reasonably well,
except for low reaction rates and fast diffusion. MF overestimates the
characteristic length $\Delta$ and allows $\rm A$'s to penetrate farther into the
pipe than in the simulations. The reason for this is that MF describes the
fact that the particles cannot pass each other by reducing the diffusion, but
this effectively does allow for passing. The larger $\Delta$ in MF means
also a larger $Q_{\rm A}$. As a consequence the $\rm B$ production in MF is larger and,
because these $\rm B$'s have to be able to leave the pipe via desorption, the
probabilities $\langle {\rm B}_1 \rangle$ and $\langle{\rm B}_S \rangle$ are larger in
MF. The probabilities $\langle {\rm A}_1 \rangle$ and $\langle {\rm A}_S\rangle$ are
therefore smaller, which means that the MF curves and the simulation curves
in figure  cross each other, as can actually be seen. The behavior of the
system at high loading and at low loading is about the same, except that
$\Delta$ is smaller at high loading.
\\
 One might expect that the larger the number of reactive sites the more $\rm B$'s
will be produced in the pipe.
 From the simulations we see that the amount of $\rm B$'s produced per unit time by all
reactive sites goes to a limit value when the number of reactive
sites is increased. In figure~\ref{Fig4}, the marked line
represents the $\rm B$ production as a function on the length of
the pipe and the dashed line the $\rm B$ production according
to MF. For short pipe lengths, the $\rm B$ production from both MF
and simulations increase linearly with $\textsl S$, while for
higher lengths it converges to a limiting value. The limiting
value is higher for MF.
 This could be seen also from the Table~\ref{tab:Table2}. According to MF there are more
$\rm B$ produced in the pipe.

 For the case $W_{\rm ads}\to\infty$ we have
\begin{equation}
{B}_{prod}={{2W_{\rm rx}W_{\rm des}}\over{W_{\rm rx}+W_{\rm des}}}.
\end{equation}
 From the simulations (see figure~\ref{Fign3}) we see that for high adsorption rates,
${B}_{prod}$ converges to a point and the corresponding value is equal to
the analytical value for the case adsorption is infinitely fast. The reason for
this is that all the sites are occupied, diffusion is completely
suppressed, and only the marginal sites play a role. The expression above
can be seen as a factor 2 for the two marginal sites, the probability that
an $\rm A$ at the marginal sites is converted to a $\rm B$ before it desorbs
$W_{\rm rx}/{W_{\rm rx}+W_{\rm des}}$, and the rate constant for desorption
$W_{\rm des}$.

 The accuracy of the simulation results for $Q_{\rm A}$ and $B_{prod}$ can be
derived by looking at the total loading $Q$ in Table~\ref{tab:Table2}. For
the total loading $Q$, the simulation results can be compared with the
values of the exact expression (12). We remark that the largest deviation
from the exact analytical results is 0.04, so the relative errors are around 0.02$\%$.

  The differences between MF and the simulations becomes especially clear in
the limit $W_{\rm diff}\to\infty$. Because this limit makes the system homogeneous
in MF we get
\begin{equation}
 Q_{\rm B}= {W_{\rm ads}\over{W_{\rm ads}+W_{\rm des}}}{W_{\rm rx}\over{W_{\rm rx}+W_{\rm des}}},
\end{equation}
\begin{equation}
 Q_{\rm A}= {W_{\rm ads}\over{W_{\rm ads}+W_{\rm des}}}{W_{\rm des}\over{W_{\rm rx}+W_{\rm des}}}.
\end{equation}
\\

\begin {figure*}
\centering
\subfigure {\epsfig {figure=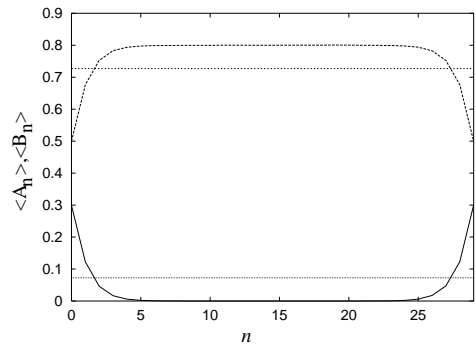, width=6.5cm} }
\caption { Analytical and simulation results for site occupancy of a
           system when parameters are: $S=N_{reac}=30$, $W_{\rm ads}$=0.8,
           $W_{\rm des}$=0.2, $W_{\rm diff}$=100 and $W_{\rm rx}$=2.
           The continuous line and the corresponding symmetric line
           represent the simulation profiles for site occupancy with $A$ and
           $B$ particles. The bottom and the upper straight lines represent the analytical
           results for occupancy with $\rm{A}$ and with $\rm {B}$ particles respectively.
         }
\label{Fig3}
\end {figure*}

 The first factor in these expressions is the probability that a site is
occupied. The second factor indicates if the particle is converted
to a $\rm B$ or not before it desorbs. The simulations show that
the system should not be homogeneous at all (see
figure~\ref{Fig3}). The $\rm B$ production increases linearly with $S$ only for
the case of infinitely-fast diffusion, otherwise it converges to a limiting
value.  

\subsection {\label{sec:lev4} Only some of the sites reactive }

We consider now the situation that not all the sites are reactive, and that these
reactive sites can be either uniformly distributed inside the pipe or
distributed in compact blocks.
 We will show that the number of reactive sites  doesn't change
qualitatively the properties of the system. $Q_{\rm A}$, $Q_{\rm B}$ and number of
$\rm B$ produced for a variable number of reactive sites are compared with the previous results.


\subsubsection { Mean Field }

From the Master Equation it is easy to show that the total loading is again just the
same as in the case when all the sites are reactive. We introduce an extra coefficient $\Delta_n$
in the MF equations to the reaction term. $\Delta_n=1$ if $n$ is a
reactive site and $\Delta_n=0$ if it is not a reactive site. The
steady-state equations are identical to equations (25), except that $W_{\rm rx}$
should be replaced by $W_{\rm rx}\Delta_n$.
The resulting set of equations is linear again and it should be possible
to solve them numerically. In fact only the
probabilities for the marginal and reactive sites have to be solved numerically.
 For the other sites the probabilities can be obtained by simple linear
interpolation. That this is correct can be seen because those sites only have
the diffusion term. We can also remove the probabilities for the $\rm B$'s, because
we have from the model without conversion that
\begin{equation}
  \expec{{\rm A}_n}+\expec{{\rm B}_n}
  =1-\expec{*_n}
  ={W_{\rm ads}\over W_{\rm ads}+W_{\rm des}}.
\end{equation}
 The resulting equations for the reactive sites have the same form as
equation (25) for the non-marginal sites. We expect therefore that we get an
exponential decrease of $\langle {\rm A}_n\rangle$ on the reactive sites when we
move from the marginal sites to the center of the pipe, and a linear
dependence on $n$ between the unreactive sites.

\subsubsection { Simulation results}

   The number of reactive sites is considered to vary from 1 to 50$\%$ and
the reactive sites are distributed either in blocks situated near the
marginal sites, in the middle of the pipe, or homogeneously distributed in
the pipe.
   We will first compare the MF results with the MC simulation results for
different sets of parameters and then we look at the dependence of $\rm B$
production and total loading $Q_{\rm A}$ on the number and position of reactive
sites.
 For the comparison between MF and MC results we consider the system size
$\textsl{S}=30$ and the number of reactive sites $N_{reac}=10$.
 The sets of parameters used for the specific situations to be studied are
the same as the sets used in the case with all the sites reactive in the
previous section.

\begin{table*}
\begin{center}
\begin{tabular}{|l| p{1.5cm} p{1.5cm} p{1.5cm} p{1.5cm} p{1.5cm} p{1.5cm} |}
\hline \multicolumn {1}{|c}{}  &\multicolumn {2}{c}{Marginal} & \multicolumn {2}{c}{Middle}& \multicolumn {2}{c|} {Homogeneous}\\
\hline  set & MF & Sim & MF & Sim & MF & Sim\\
\hline a)& 0.0512    & 0.0500   & 0.0731  & 0.0771 & 0.0208  & 0.0469 \\
 b)& 0.0153    & 0.0152   & 0.0712  & 0.0788 & 0.0138  & 0.0206\\
 c)& 0.0590    & 0.0672   & 0.0901  & 0.0881 & 0.0719  & 0.0594\\
 d)& 0.0041    & 0.0041   & 0.0667  & 0.0730 & 0.0120  & 0.0123\\
 e)& 0.0067    & 0.0006   & 0.0447  & 0.0583 & 0.0126  & 0.0121\\
 f)& 0.0896    & 0.0752   & 0.3008  & 0.3473 & 0.1121  & 0.1056 \\
 g)& 0.0376    & 0.0369   & 0.2850  & 0.3383 & 0.0585  & 0.0605\\
 h)& 0.0871    & 0.0579   & 0.2844  & 0.3250 & 0.1227  & 0.0867\\
 i)& 0.0048    & 0.0048   & 0.2606  & 0.3137 & 0.0319  & 0.0413\\
 j)& 0.0056    & 0.0053   & 0.1556  & 0.2826 & 0.0175  & 0.0289\\
\hline
\end{tabular}
\end{center}
\caption{\label{tab:table3}  Simulation and MF results for $Q_{\rm
A}$  for all the sets of parameters in the cases of homogeneous
distribution of the reactive sites, blocks of reactive sites in
the middle of the pipe and near the marginal sites ($S=30$,
$N_{reac}=10$)}
\end{table*}

\begin{table*}
\begin{center}
\begin{tabular}{|l| p{1.5cm} p{1.5cm} p{1.5cm} p{1.5cm} p{1.5cm} p{1.5cm} |}
\hline \multicolumn {1}{|c}{}  &\multicolumn {2}{c}{Marginal} & \multicolumn {2}{c}{Middle}& \multicolumn {2}{c|} {Homogeneous}\\
\hline  set & MF & Sim & MF & Sim & MF & Sim\\
\hline a)& 0.0099    & 0.0099    & 0.0001 & 0.0000 & 0.0011 & 0.0045\\
 b)& 0.0449    & 0.0477    & 0.0001 & 0.0008 & 0.0153 & 0.0117\\
 c)& 0.1156    & 0.1021    & 0.0393 & 0.0216 & 0.0728 & 0.0521\\
 d)& 0.2449    & 0.2492    & 0.0283 & 0.0161 & 0.1357 & 0.0891\\
 e)& 0.2767    & 0.2763    & 0.1482 & 0.0646 & 0.2410 & 0.1899\\
 f)& 0.0239    & 0.0235    & 0.0014 & 0.0006 & 0.0086 & 0.0076\\
 g)& 0.1129    & 0.1160    & 0.0015 & 0.0000 & 0.0139 & 0.1171\\
 h)& 0.1796    & 0.1421    & 0.0470 & 0.0059 & 0.1212 & 0.0661\\
 i)& 0.2931    & 0.2941    & 0.0288 & 0.0069 & 0.1526 & 0.0897\\
 j)& 0.3013    & 0.2965    & 0.1552 & 0.0143 & 0.2713 & 0.1739\\
\hline
\end{tabular}
\end{center}
\caption{\label{tab:table4} Simulation and MF results for
${B}_{prod}$  for all the sets of parameters in the cases of
homogeneous distribution of the reactive sites, blocks of reactive
sites in the middle of the pipe and near the marginal sites
($S=30$, $N_{reac}=10$)}
\end{table*}

  We can see from the Tables~\ref{tab:table3} and ~\ref{tab:table4} that when the reactive sites are homogeneously
distributed or situated as a block in the middle of the pipe, there are
significant differences between MF results and MC results.
 When the reactive sites form blocks near the marginal sites, the results
are almost the same as when all sites are reactive: the MC and the MF
results differ if we have fast diffusion and slow reaction. The sites in the
center of the pipe are not relevant when the sites at the ends of the pipe
are reactive.
  When the reactive sites are situated only in the middle of the pipe, we have
deviations for all the sets of parameters.
They are very prominent for the case when we have high loading, fast
diffusion and fast reaction. MF strongly underestimates $\rm A$'s for all non-reactive sites,
but we have also important deviations for high loading in the cases with
fast diffusion-slow reaction, slow diffusion-fast reaction, and slow diffusion-slow
reaction. This is happening because for high loading, the end sites will
always be occupied by a particle $\rm A$ and the $\rm B$'s will not be able to get out
of the pipe. In MF particles effectively can pass each other, so $\rm B$ particles
are then able to get out of the pipe.
 Even for the case of low loading we still have deviations from MF for fast
diffusion and fast reaction. In this case MF overestimates $\rm A$'s for nonreactive
sites. For fast diffusion and slow reaction, MF underestimates $\rm A$'s for nonreactive
sites and for slow diffusion and slow reaction MF overestimates $\rm B$'s for the
reactive sites in the middle.
\begin {figure*}
\centering

\subfigure {\epsfig {figure=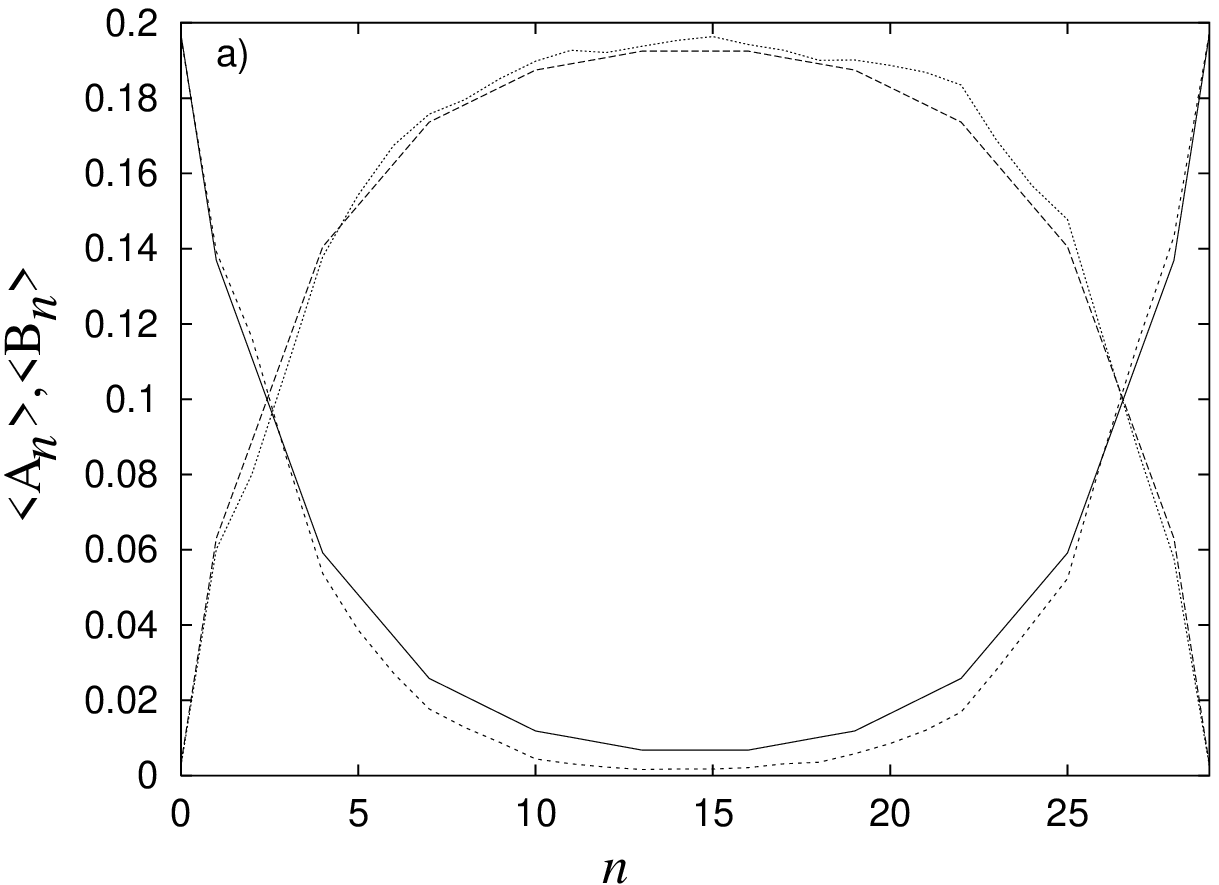, width=4.0cm} }
\subfigure {\epsfig {figure=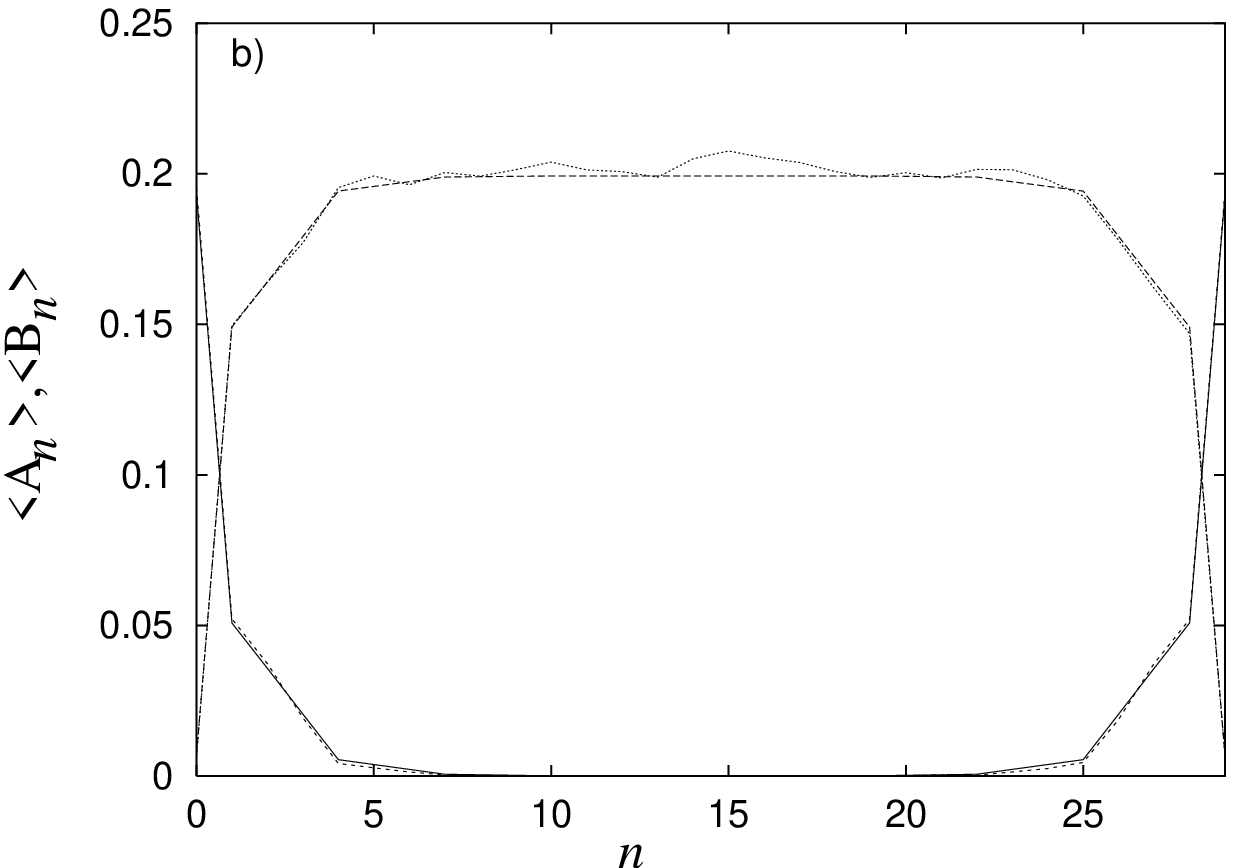, width=4.0cm} }
\subfigure {\epsfig {figure=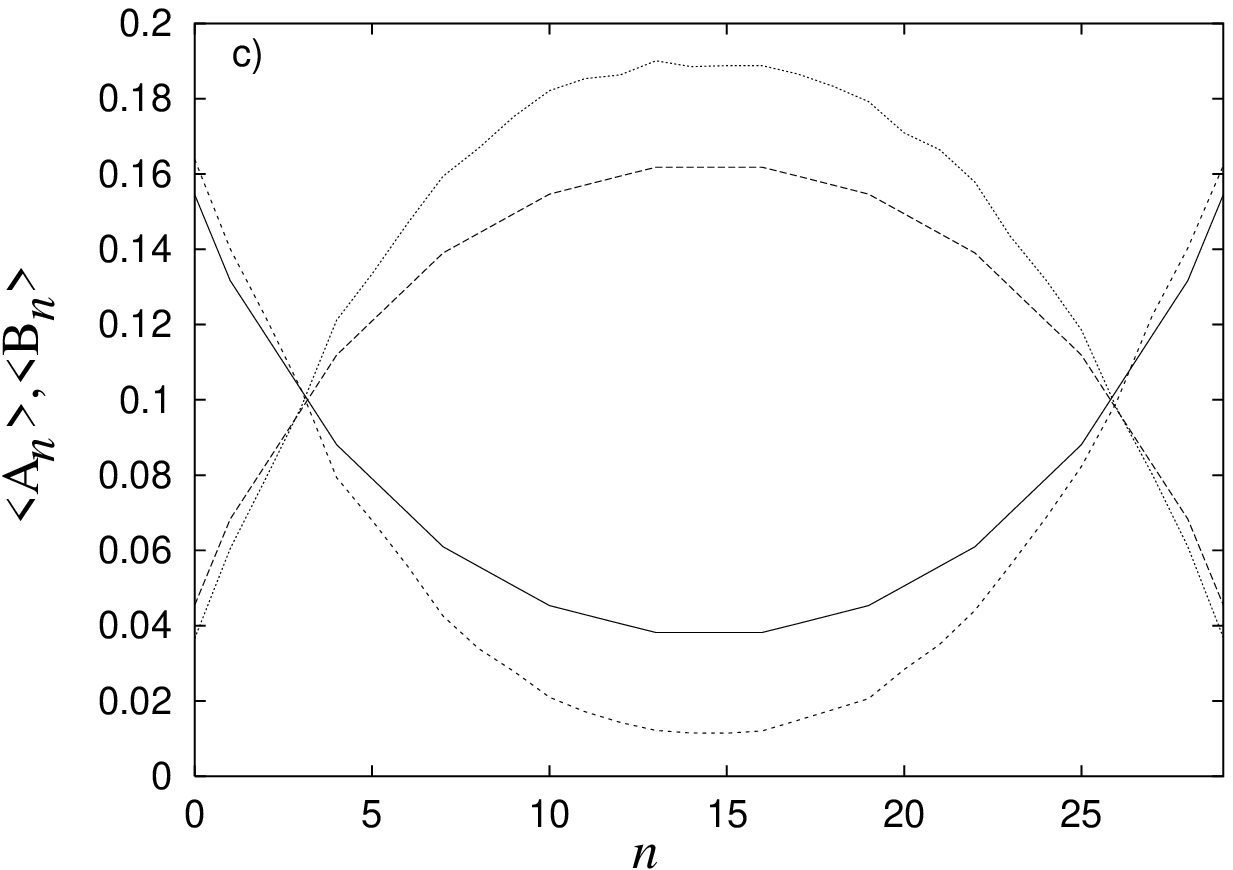, width=4.0cm} }
\subfigure {\epsfig {figure=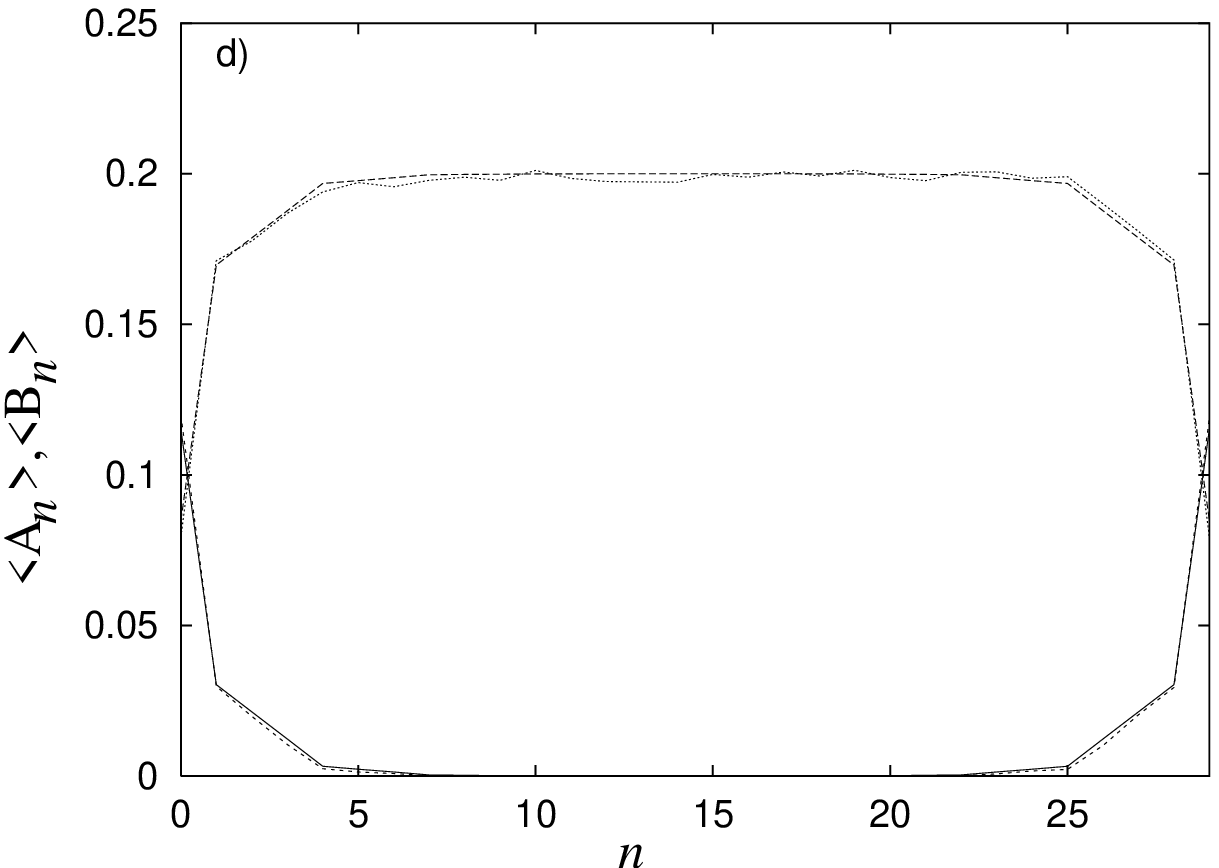, width=4.0cm} }
\subfigure {\epsfig {figure=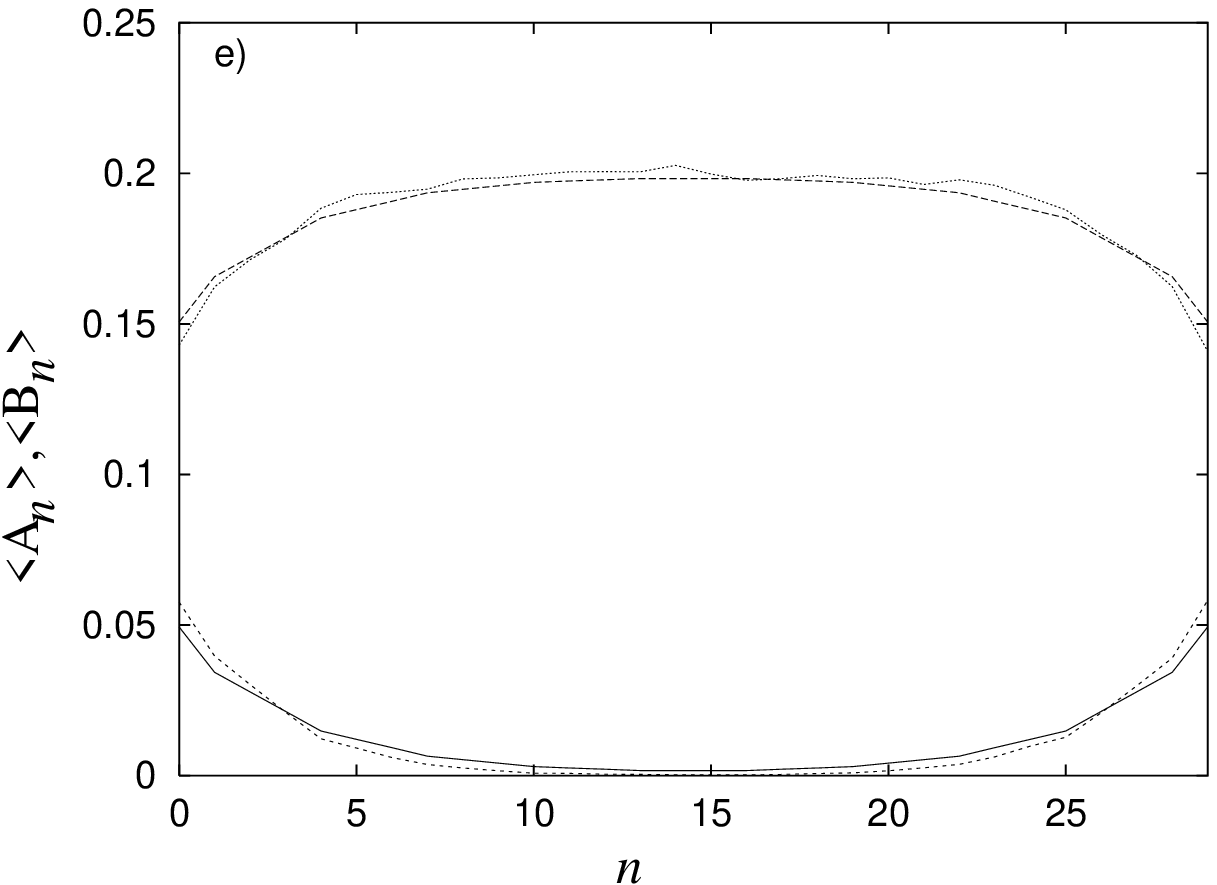, width=4.0cm} }
\caption { The site occupancy for the cases a, b, c, d, e
-homogeneous distribution. The continuous line and the
corresponding symmetric line represent the MF results. The other
dashed lines represent the DMC results. }
\label{Figurinel}
\end {figure*}

\begin {figure*}
\centering


\subfigure {\epsfig {figure=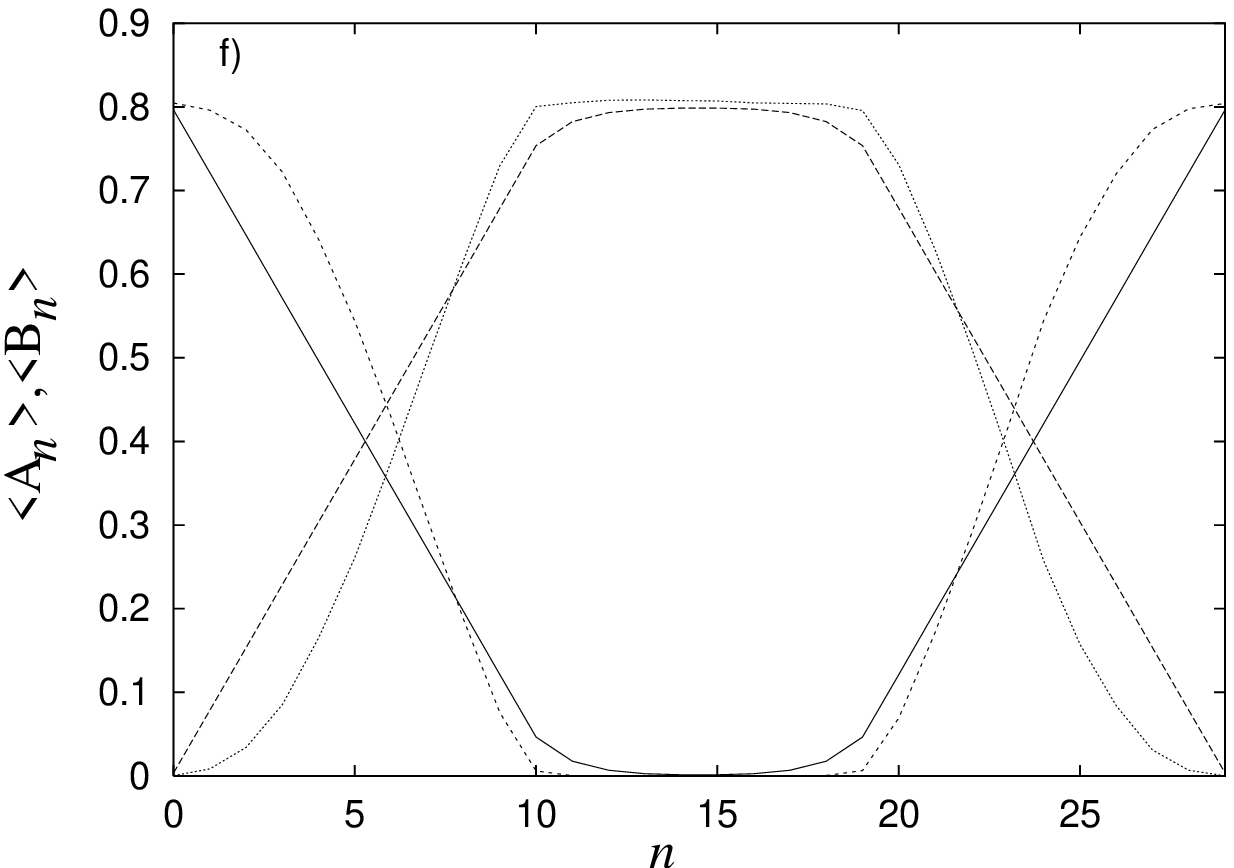, width=4.0cm} }
\subfigure {\epsfig {figure=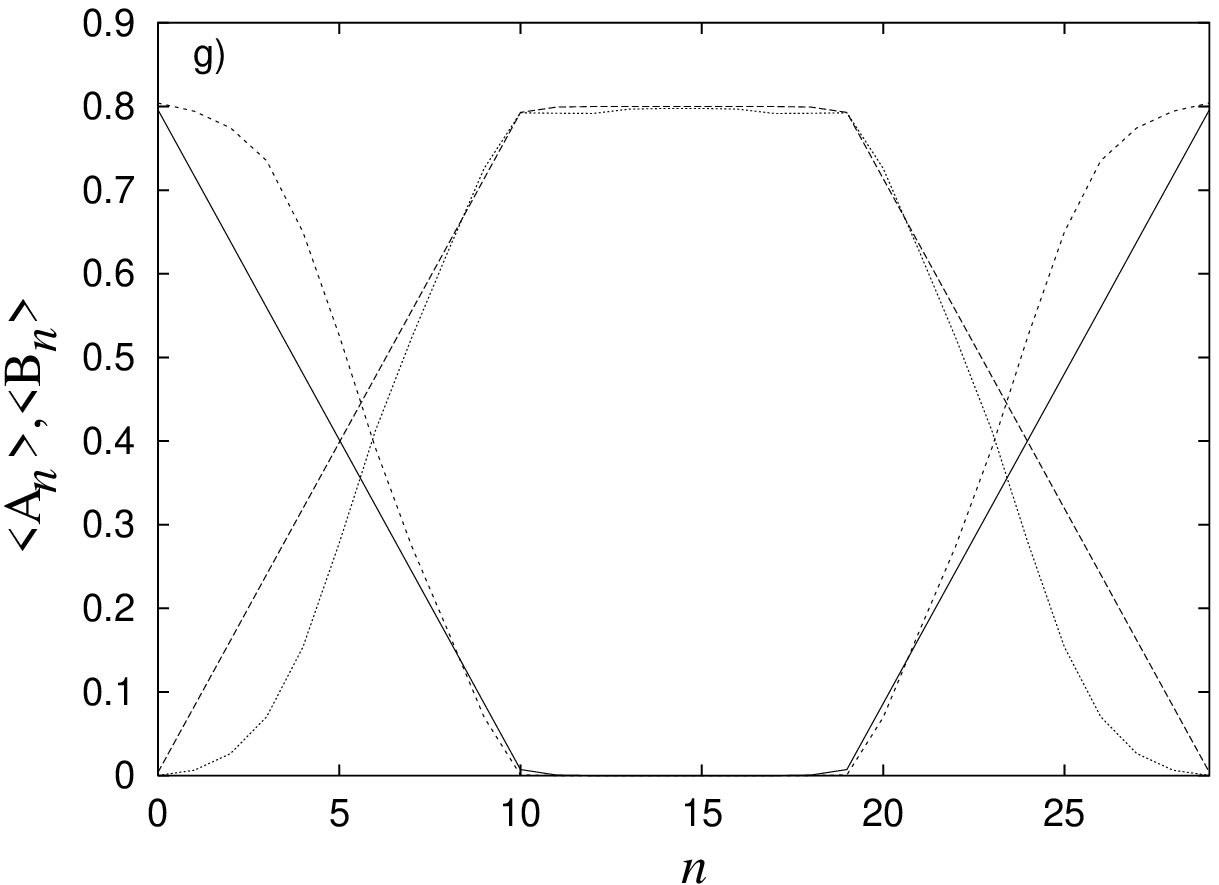, width=4.0cm} } 
\subfigure {\epsfig {figure=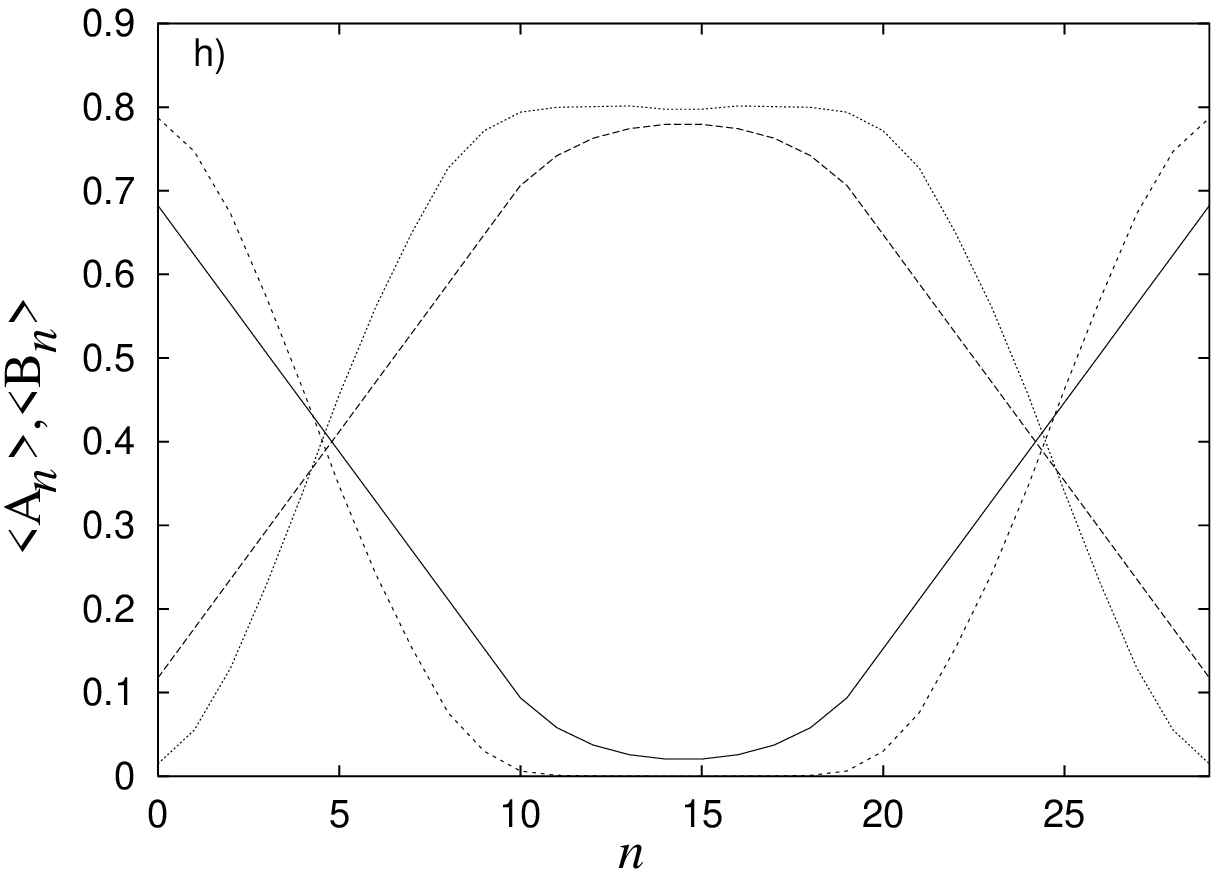, width=4.0cm} }
\subfigure {\epsfig {figure=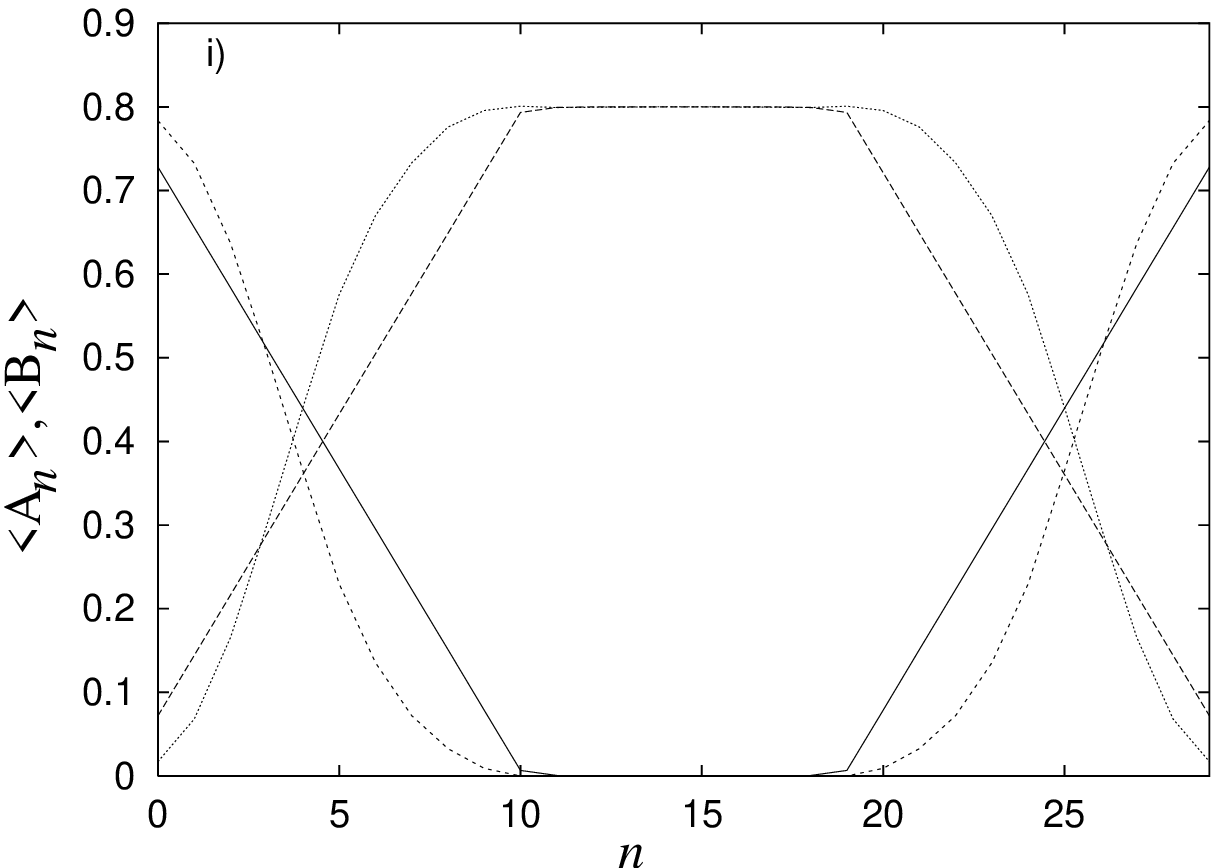, width=4.0cm} }
\subfigure {\epsfig {figure=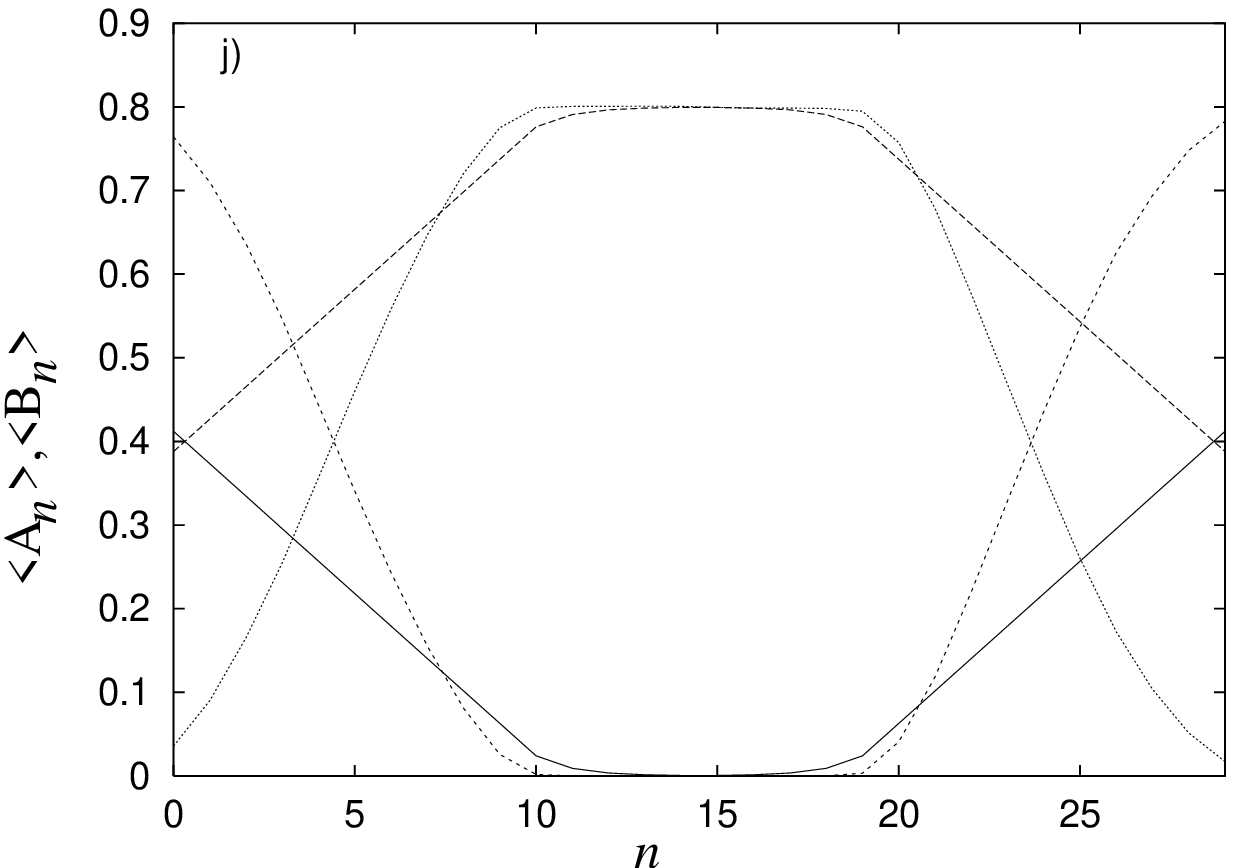, width=4.0cm} }
\caption { The site occupancy
for the cases f, g, h, i, j - middle sites reactive,
$N_{reac}=10$. The continuous line and the corresponding symmetric
line represent the MF results. The other dashed lines represent
the DMC results.}
\label{Figurinel1}
\end {figure*}


 Figures ~\ref{Figurinel} and ~\ref{Figurinel1} show how the probabilities $\langle{{\rm A}_n}\rangle$ and
$\langle{ {\rm B}_n }\rangle$ vary in the pipe. The situations for
reactive sites forming blocks at the ends of the pipe are not
shown as they are almost the same as when all the sites are
reactive (see figure 4). When the reactive sites are homogeneously
distributed the plots look also very similar to the ones with all
sites reactive, except that the characteristic length $\Delta$ is
larger.
 $\langle{{\rm A}_n}\rangle$ and $\langle {{\rm B}_n}\rangle$ look very different when
the reactive sites form a block in the middle of the pipe. The MF results
show, as predicted, a linear behavior at the nonreactive sites. The
MC results show, however, a nonlinear behavior in the form of S-like curves.
 At the reactive sites the behavior is similar to the situation with all
sites reactive with the MC results showing a more rapid approach to the
value at the middle of the pipe than MF, i.e., smaller $\Delta$. The values
at the marginal sites can differ between MC and MF quite a lot.
  This reflects the difference in ${B}_{prod}$ mentioned before: a different
${B}_{prod}$ must be accompanied by a different $\rm B$ desorption at steady-state.
  As we have already seen from the case when all the sites were reactive,
${B}_{prod}$ very rapidly approaches the limiting value when the
pipe is made longer (see figure 5). Similarly when we start with
few reactive sites and, instead of increasing the length of the
pipe, we increase the number of reactive sites. The loading
$Q_{\rm B}$ is already almost the same as the value with all sites
reactive when only about 10$\%$ of all sites are reactive provided
there are reactive sites at or very near the marginal sites. If
the reactive sites are moved away from the ends of the pipe, then
the loading $Q_{\rm B}$ and the $\rm B$ production decreases.


\section { Summary }

  We have used analytical and  simulation techniques to study the reactivity
in Single-File Systems.

  The MF results show that MF models Single-File behavior by
changing the diffusion  rate constant, but it effectively does allow passing of particles.

  When all the sites are reactive, the simulation and MF results are very
similar for all the parameters, except for the case when we have low reaction rates and
fast diffusion. In these cases MF overestimates
the amount of $\rm A$'s in the pipe. The amount of $\rm B$ produced per
unit time by all reactive sites goes to a limit value when the number of
reactive sites is increased. For high adsorption rates, ${B}_{prod}$ converges
to a point and the corresponding value is equal to the analytical value for
the case adsorption is infinitely fast. The sites in the middle of the pipe
have no effect on the $B$ production. The differences between MF and the
simulations becomes especially clear in the limit ${W_{\rm diff}}\to\infty$.

 When only some of the sites are reactive, there are significant differences
between MF and MC results when the reactive sites are homogeneously
distributed or situated as a block in the middle of the pipe.
 When the reactive sites form blocks near the marginal sites, the results
are almost the same as when all sites are reactive: The MC and the MF
results differ only when we have fast diffusion and slow reaction. The sites in the
center of the pipe are not relevant when the sites at the ends of the pipe
are reactive.
  When the reactive sites are situated in the middle of the pipe, we have
deviations for all the sets of parameters.
They are very prominent for the case when we have high loading, fast
diffusion and fast reaction. MF strongly underestimates $\rm A$'s for all non-reactive sites,
but we have also important deviations for high loading in the cases with
fast diffusion-slow reaction, slow diffusion-fast reaction, slow diffusion-slow
reaction.
 The MF results show a linear behavior at the nonreactive sites. The
MC results show, however, a nonlinear behavior in the form of S-like curves.
 The loading $Q_{\rm B}$ is already almost the same as the value
with all sites reactive when only about 10$\%$ of all sites are reactive
provided there are reactive sites at or very near the marginal sites. If the
reactive sites are moved away from the ends of the pipe, then the loading
$Q_{\rm B}$ and the $\rm B$ production decreases.

\section {Acknowledgments}

 The authors thank Prof.dr. R.A. van Santen for many stimulating
discussions.


\section {Appendix}

\subsection {Probability to find the system in a certain
configuration. Loadings and fluctuations.}

   We show the existence of a function $q$, depending only on the number of
particles such that \begin{equation}
  P_\alpha=q\big(n(\alpha)\big)
\end{equation} is the steady-state solution of the Master Equation (2) for a
system without conversion, where $n(\alpha)$ is the number of particles in configuration $\alpha$.
The second part of the proof consists of showing the uniqueness of the solution.

 Substitution of $P_\alpha=q\big(n(\alpha)\big)$ in equation (9) shows that the
last term in the Master Equation vanishes, because
$\Delta_{\alpha\beta}^{({\rm diff})}=\Delta_{\beta\alpha}^{({\rm diff})}$
and $n(\alpha)=n(\beta)$. The other terms can also
be simplified by using how the number of particles changes upon
adsorption and desorption.
\begin{align}
  {dP_\alpha\over dt}
  &=W_{\rm ads}\left[
    q\big(n(\alpha)-1\big)\sum_\beta\Delta_{\alpha\beta}^{({\rm ads})}
   -q\big(n(\alpha)\big)\sum_\beta\Delta_{\beta\alpha}^{({\rm ads})}
    \right] \nonumber \\
  &+W_{\rm des}\left[
    q\big(n(\alpha)+1\big)\sum_\beta\Delta_{\alpha\beta}^{({\rm des})}
   -q\big(n(\alpha)\big)\sum_\beta\Delta_{\beta\alpha}^{({\rm des})}
    \right].
\end{align}
A further simplification is possible if we realize that desorption
reverses the effect of an adsorption and {\it vice versa}. This means
$\Delta_{\alpha\beta}^{({\rm des})}=\Delta_{\beta\alpha}^{({\rm ads})}$. This leads to
\begin{align}
  {dP_\alpha\over dt}
  &=\left[
     q\big(n(\alpha)-1\big)W_{\rm ads}
   - q\big(n(\alpha)\big)W_{\rm des}
    \right]
    \sum_\beta\Delta_{\alpha\beta}^{({\rm ads})}\nonumber \\
  &-\left[
     q\big(n(\alpha)\big)W_{\rm ads}
   - q\big(n(\alpha)+1\big)W_{\rm des}
    \right]
    \sum_\beta\Delta_{\beta\alpha}^{({\rm ads})}.
\end{align}
We denote by $N$, the number of particles in a certain configuration,
$N=n(\alpha)$.
We see that we get a steady-state solution for \begin{equation} {dP_\alpha\over
dt}=0, \end{equation} provided by
\begin{equation}
  {q(N+1)\over q(N)}={W_{\rm ads}\over W_{\rm des}}
\end{equation}
for $N={\textsl{0, 1, 2,\ldots, S-1}}$. (Note that the case $N=S$ in the Master
Equation presents no problems, because the summation over $\beta$ yields zero.)

The second step consists of showing that this solution is the only one.
This part for instance can be found in Chapter~5 of Van Kampen.~\cite{kampen}

\subsection {Derivation of function q(N)}

 Expression (45) leads to
\begin{equation}
  q(N)=C\left[{W_{\rm ads}\over W_{\rm des}}\right]^N,
\end{equation}
where $C$ is some normalization constant. We can compute it from
\begin{align}
  1&=\sum_\alpha P_\alpha
    =\sum_\alpha q\big(n(\alpha)\big)
    =\sum_{N=0}^S\left({S\atop N}\right)q(N) \nonumber \\
   &=C\sum_{N=0}^S\left({S\atop N}\right)
     \left[{W_{\rm ads}\over W_{\rm des}}\right]^N
    =C\left[{W_{\rm des}+W_{\rm ads}\over W_{\rm des}}\right]^S.
\end{align}
The combinatorial factor after the third equal sign derives from the
number of configurations with $N$ particles. The last step uses
\begin{equation}
  (x+y)^S=\sum_{N=0}^S\left({N\atop n}\right)x^{N-n}y^n.
\end{equation}
The expression for $q(N)$ now becomes
\begin{equation}
  q(N)=\left[{W_{\rm des}\over W_{\rm des}+W_{\rm ads}}\right]^S
       \left[{W_{\rm ads}\over W_{\rm des}}\right]^N.
\end{equation}
Note that this expression does not depend on $W_{\rm diff}$: i.e.,
diffusion has no effect at all on steady-state properties.

 The probability $p(N)$ that there are $N$ particles in the system is given by

\begin{equation}
  p(N)=\begin{pmatrix}
         S \cr
         N \cr
       \end{pmatrix}q(N)
      =\left({S\atop N}\right)
       \left[{W_{\rm des}\over W_{\rm des}+W_{\rm ads}}\right]^S
       \left[{W_{\rm ads}\over W_{\rm des}}\right]^N.
\end{equation}
This follows from (49). With this formula we can compute all statistical
properties of the number of particles. The average number of particles is
\begin{align}
  \sum_{N=0}^SN\,p(N)
  &=\left[{W_{\rm des}\over W_{\rm des}+W_{\rm ads}}\right]^S
    \sum_{N=0}^S
    \left({S\atop N}\right)N
    \left[{W_{\rm ads}\over W_{\rm des}}\right]^N\cr
  &={W_{\rm ads}\over W_{\rm des}+W_{\rm ads}}S.
\end{align}
The loading of the pipe, defined as the average number of particles per
site, is
\begin{equation}
  {Q_{\rm A}={{\sum_{N=0}^SN\,p(N)}\over S } }={W_{\rm ads}\over W_{\rm ads}+W_{\rm des}}.
\end{equation}
The average squared number of particles is
\begin{align}
  \sum_{N=0}^SN^2\,p(N)
  &=\left[{W_{\rm des}\over W_{\rm des}+W_{\rm ads}}\right]^S
    \sum_{N=0}^S
    \left({S\atop N}\right)N^2
    \left[{W_{\rm ads}\over W_{\rm des}}\right]^N\cr
  &={W_{\rm ads}(W_{\rm des}+SW_{\rm ads})
     \over(W_{\rm des}+W_{\rm ads})^2}S.
\end{align}
The variance, i.e., the square of the fluctuation in the number of
particles, is then
\begin{equation}
  \sum_{N=0}^SN^2\,p(N)-\left[\sum_{N=0}^SN\,p(N)\right]^2
  ={W_{\rm ads}W_{\rm des}\over(W_{\rm des}+W_{\rm ads})^2}S.
\end{equation}

\subsection {Derivation of the one-site and two-sites occupancy for the model
without conversion}

The probability that site $n$ is occupied by $\rm A$ is given by
\begin{equation}
\begin{split}
{\langle {\rm A}_n \rangle}
 &=\sum_{\alpha}P_{\alpha}{\Delta}_\alpha^{(n)},\cr
 &=\sum_N\sum_{\alpha \in N}P_{\alpha}{\Delta}_\alpha^{(n)},\cr
 &=\sum_{N} q(N) \sum_{\alpha \in N}{\Delta}_\alpha^{(n)},\cr
 &=\sum_{N=1}^Sq(N)\left({{S-1}\atop {N-1}}\right),\cr
 &={\left[W_{\rm des}\over{W_{\rm des}+W_{\rm ads}}\right]}^S \sum_{N=1}^S \left({{S-1}\atop{N-1}}\right){\left[{W_{\rm ads}}\over{W_{\rm des}}\right]}^N,\cr
 &={\left[ W_{\rm des}\over{W_{\rm des}+W_{\rm ads}} \right]}^S {W_{\rm ads}\over{W_{\rm des}}} {\left[1+{W_{\rm ads}\over{W_{\rm des}}}\right]}^{S-1},\cr
 &={W_{\rm ads}\over{W_{\rm ads}+W_{\rm des}}},\cr
\end{split}
\end{equation}
where ${\Delta}_\alpha^{(n)}$ is 1 if site $n$ in configuration
$\alpha$ is occupied by an $\rm A$ particle, and it is 0
otherwise. The combinatorial factor denotes the number of ways the
particles except the one at site $n$ can be distributed over the
remaining sites.
 Knowing the one-site occupancy we can derive the two-site occupancy

\begin{equation}
\begin{split}
{\langle {\rm A}_n {\rm A}_{n+1}\rangle}
 &=\sum_{\alpha}P_{\alpha}\Delta_\alpha^{(n)}\Delta_\alpha^{(n+1)},\cr
 &=\sum_N\sum_{\alpha \in N}P_{\alpha}\Delta_\alpha^{(n)}\Delta_\alpha^{(n+1)},\cr
 &=\sum_{N} q(N) \sum_{\alpha \in N}\Delta_\alpha^{(n)}\Delta_\alpha^{(n+1)},\cr
 &=\sum_{N=2}^Sq(N)\left({{S-2}\atop {N-2}}\right),\cr
 &={\left[W_{\rm des}\over{W_{\rm des}+W_{\rm ads}}\right]}^S \sum_{N=2}^S \left({{S-2}\atop{N-2}}\right){\left[{W_{\rm ads}}\over{W_{\rm des}}\right]}^N,\cr
 &={\left[ W_{\rm des}\over{W_{\rm des}+W_{\rm ads}} \right]}^S {\left({W_{\rm ads}\over{W_{\rm des}}}
\right )}^2 {\left[1+{W_{\rm ads}\over{W_{\rm des}}}\right]}^{S-2},\cr
 &={\left({W_{\rm ads}\over{W_{\rm ads}+W_{\rm des}}}\right)}^2.\cr
\end{split}
\end{equation}


\subsection{Continuum limit}

  The rate equation for the A's is
\begin{equation}
\begin{split}
{d\langle {\rm A}_n \rangle }\over{dt} &={W}_{\rm diff}[{\langle
{\rm A}_{n-1} *_n \rangle}+ {\langle *_n {\rm A}_{n+1}
\rangle}-{\langle {\rm A}_n *_{n+1} \rangle}\cr
&-{\langle*_{n-1}{\rm A}_n \rangle}]-W_{\rm rx}{\langle {\rm
A}_n\rangle}.\cr
\end{split}
\end{equation}
The MF approximation of this equation is
\begin{equation}
\begin{split}
{d\langle {\rm A}_n \rangle }\over{dt} &={W}_{\rm
diff}[({\langle{\rm A}_{n-1} \rangle}+{\langle {\rm A}_{n+1}
\rangle}) {\langle *_n \rangle}\cr
 &-{\langle {\rm A}_n
\rangle}(\langle *_{n-1} \rangle + \langle *_{n+1} \rangle)]\cr
 &-W_{\rm rx}{\langle {\rm A}_n \rangle}.\cr
\end{split}
\end{equation}
If we take the continuum limit and denote by $a=a(x,t)$,
$b=b(x,t)$ and $v=v(x,t)$ the probability distribution of $\rm
A$'s, $\rm B$'s and vacancies respectively, and if we use Taylor
series for the diffusion term, the equation becomes

\begin{widetext}
\begin{equation}
\begin{split}
{{\partial a}\over{\partial t}} &={W}_{\rm
diff}[a((n-1)d,t)+a((n+1)d,t)]v(nd,t)\cr
   & - {W}_{\rm diff}a(nd,t)[v((n-1)d,t)+v((n+1)d,t)]a(nd,t) - W_{\rm rx}a \cr
&\approx {W}_{\rm diff} \left[a-d{{\partial a}\over{\partial
x}}+{1\over{2}}d^2{{\partial^2 a}\over{\partial
   x^2}} + a + d {{\partial a}\over{\partial x}} + {1\over{2}} {{\partial^2 a}\over{\partial x^2}}\right]v \cr
   & - {W}_{\rm diff} \left[v-d{{\partial v}\over{\partial x}}+{1\over{2}}d^2{{\partial^2 v}\over{\partial
  x^2}} + v + d {{\partial v}\over{\partial x}} + {1\over{2}} {{\partial^2 v}\over{\partial x^2}}\right]a -W_{\rm rx}a \cr
&={W}_{\rm diff}\left[2a + d^2 {{\partial^2 a}\over{\partial
x^2}}\right]v - {W}_{\rm diff}a\left[2v+d^2 {{\partial^2
v}\over{\partial x^2}}\right] - W_{\rm rx}a \cr &={W}_{\rm
diff}d^2\left[v {{\partial^2 a}\over{\partial x^2}} + a
{{\partial^2 a}\over{\partial x^2}} + a {{\partial^2
b}\over{\partial x^2}}\right] - W_{\rm rx}a \cr &={W}_{\rm
diff}d^2\left[(1-b){{\partial^2 a}\over{\partial
x^2}}+a{{\partial^2 b}\over{\partial x^2}}\right] - W_{\rm
rx}a,\cr
\end {split}
\end {equation}
\end{widetext}
where $d$ is the distance between sites. A similar relation can be
derived for $b(x,t)$. With $D\equiv {W_{\rm diff}}d^2$ we can
write
\begin{widetext}
\begin{equation}
 \begin{pmatrix}
    {\partial a}/{\partial t}\cr
    {\partial b}/{\partial t}\cr
 \end{pmatrix}
 =D\begin{pmatrix}
   1-b & a \cr
   b& 1-a \cr
   \end{pmatrix}
   \begin{pmatrix}
    {\partial^2 a }/\partial x^2\cr
    {\partial^2 b}/\partial x^2\cr
   \end{pmatrix}
  + W_{\rm rx}\begin{pmatrix}
   -a \cr
    a \cr
    \end{pmatrix}.
\end{equation}
\end{widetext}

\subsection { MF derivation of the total loading $Q_A$ in case with conversion}
 The total loading with $\rm A$'s, $Q_{\rm A}$, is written as
\begin{equation}
 Q_{\rm A}={1\over{S}}\sum_{n=1}^S \langle {\rm A}_n\rangle,
\end{equation}
so, the expression for $Q_{\rm A}$ is
\begin{equation}
\begin{split}
Q_{\rm A}
    &={a\over{S}}\sum_{n=1}^S[{x_1}^n + {x_1}^{S+1-n}],\cr
    &={a\over{S}}\sum_{n=1}^S{x_1}^n + {a\over{S}}\sum_{n=1}^S{x_1}^{S+1-n},\cr
    &={{2a}\over{S}}\sum_{n=1}^S{x_1}^n,\cr
    &={{2a}\over{S}} {{x_1{(1-{x_1}^S)}}\over{1-x_1}}.
\end{split}
\end{equation}

\bibliographystyle{./apsrev}
\nocite{*}

\end {document}